\documentclass[chicago, iop, apj, numberedappendix]{emulateapj}

\usepackage[colorinlistoftodos]{todonotes}
\usepackage[colorlinks=true, allcolors=blue]{hyperref}
\usepackage[english]{babel}
\usepackage[utf8x]{inputenc}
\usepackage[T1]{fontenc}
\usepackage[normalem]{ulem}

\usepackage{aas_macros}
\usepackage{natbib}
\usepackage{bm}
\usepackage{times}
\usepackage{color}

\usepackage{amsmath}
\usepackage{array}
\usepackage{amsfonts}
\usepackage{graphicx}

\newcommand{\zrange}{$2 < z < 3.2$}
\newcommand{\cii}{[C{\sc\,ii}]}
\newcommand{\Lsun}{{\rm L_\odot}}

\shorttitle{Correlated continuum emission in IM}
\shortauthors{Authors}
\keywords{GRE}

\begin{document}

\author{
E.~R.~Switzer\altaffilmark{1}, C.~J.~Anderson\altaffilmark{1}, A.~R.~Pullen\altaffilmark{2}, S.~Yang\altaffilmark{2}
}

\altaffiltext{1}{NASA Goddard Space Flight Center, Greenbelt, MD 20771, USA}
\email{eric.r.switzer@nasa.gov}
\altaffiltext{2}{Center for Cosmology and Particle Physics, Department of Physics, New York University, 726 Broadway, New York, NY, 10003, USA}

\shortauthors{Switzer}

\title{INTENSITY MAPPING IN THE PRESENCE OF FOREGROUNDS AND CORRELATED CONTINUUM EMISSION \shorttitle{Intensity mapping with correlated continua}}

\begin{abstract}
Intensity mapping has attracted significant interest as an approach to measure the properties of the interstellar medium in typical galaxies at high redshift. Intensity mapping measures the statistics of surface brightness as a function of frequency, making it sensitive not only to all line emission of interest but also radiation from all other sources. Significant effort has gone into developing approaches that reject foreground contamination. Additionally, the target galaxies have multiple sources of emission that can complicate the interpretation of the line brightness. We describe the problem of jointly estimating correlated continuum emission and cleaning uncorrelated continuum emission, such as from the Milky Way. We apply these considerations to a cross-correlation of Planck data with BOSS quasars for a determination of \cii\ for \zrange. Intensity mapping surveys with few bands have unique challenges for treating foregrounds and avoiding bias from correlated continuum emission. We show how a future intensity mapping survey with many bands can separate line from continuum emission in cross-correlation.
\end{abstract} 

\section{Introduction}
\label{sec:introduction}

A major goal of modern astrophysics is to understand the evolution of galaxies in their cosmological context. Two epochal shifts of great interest are the formation of the galaxies that reionize the universe and the subsequent, dramatic decline in the star formation rate from $z{\sim}2$ to the present \citep{2014ARA&A..52..415M} despite the continued growth of dark matter halos. Studies of CO, \cii, and $21$\,cm line emission are particularly valuable as tracers of star formation and its precursors in the interstellar medium during these eras \citep{2013ARA&A..51..105C}.

Most studies of line emission to-date measure the properties of individual galaxies to draw broader conclusions about galaxy evolution. These galaxies are often selected from surveys targeting very luminous objects, such as quasars and luminous IR galaxies \citep{2013ARA&A..51..105C}. Such catalogs provide a biased sample of gas tracers rather than measuring the average population. Even when line emission is detected blindly (e.g., \citet{2016ApJ...833...69D}), these detections may be limited to the brightest objects, so are not representative of most galaxies. Blind surveys are often in cosmologically small volumes (especially at low redshift), making the source counts subject to cosmic variance \citep{2010ApJ...716L.229R}. Furthermore, surveys for individual objects are expensive because they must detect galaxies at high significance, which requires large apertures or interferometers to gain high flux sensitivity and avoid confusion. 

Intensity mapping (IM) \citep{1979MNRAS.188..791H, 1990MNRAS.247..510S,1997ApJ...475..429M,1999ApJ...512..547S,2008MNRAS.383.1195W,2008PhRvL.100i1303C,2010JCAP...11..016V, 2011JCAP...08..010V} overcomes many of these challenges, so is attracting increasing interest and investment, as detailed in the recent whitepaper \citet{2017arXiv170909066K}. Rather than identifying individual objects as in a galaxy redshift survey, IM measures the integrated emission at a given (observed) frequency from galaxies or the IGM. IM requires only modest aperture sizes to reach the smallest scales of cosmological interest, regardless of confusion. The method was originally developed for $21$\,cm radiation from reionization \citep{1979MNRAS.188..791H, 1990MNRAS.247..510S, 1997ApJ...475..429M}, but has expanded to numerous lines and science goals. Observations of CO, \cii, and $21$\,cm can constrain the luminosity function of galaxies too faint to observe individually at high redshift. These observations will be critical for constraining the most uncertain aspects of current galaxy formation models, namely, the role of star formation and ``feedback" from stars, supernovae, and active black holes. 

Intensity mapping has already provided several measurements that bear on galaxy evolution. Cross-correlation of $21$\,cm emission and WiggleZ \citep{2013MNRAS.434L..46S} has determined the HI abundance at $z{\approx}0.8$, which is consistent with feedback from active galactic nuclei \citep{2016MNRAS.456.3553V, 2017MNRAS.469.2323P}. Preliminary indication of \cii\ emission in cross-correlation between Planck and the BOSS quasar sample for \zrange\ \citep{2018MNRAS.478.1911P} favors collisional excitation models at these redshifts that are different from local scaling relations. COPSS-II \citep{2016ApJ...830...34K} observed CO to provide the first inferences of the molecular gas abundance $2.3 < z < 3.3$, ruling out several models. \citet{2018arXiv180606050C} find diffuse ${\rm Ly}\alpha$ emission in cross-correlation with the ${\rm Ly}\alpha$ forest and quasars at a level that is consistent with hydrodynamic simulations.\citet{2017arXiv170909066K} describes several dedicated IM experiments, both ongoing and future.

An interpretation of an autonomous intensity mapping survey must argue that there is no bias from residual bright continuum radiation (Milky Way and extragalactic), interloper lines from other redshifts, or instrumental response. Cross-correlation with overdensities inferred from a galaxy redshift survey provides an approach to circumvent this foreground contamination. Here, uncorrelated foreground variance increases errors but does not bias the determination. Cross-correlation also allows breakdown by galaxy type or environment to probe the impact of those factors on the ISM \citep{2017MNRAS.470.3220W}. 

We describe several challenges and approaches related to cross-correlation in the specific context of the measurement in \citet{2018MNRAS.478.1911P}. Here, the cross-power of Planck 545\,GHz and BOSS quasars is used to constrain \cii\ across \zrange. \cii\ is a tracer of great interest for intensity mapping. Star formation excites the $157.7\,\mu{\rm m}$ ($1900$\,GHz) $^2P_{3/2} \rightarrow\,^2P_{1/2}$ fine structure transition of \cii\ in several phases of the ISM. \cii\ is the brightest far-IR cooling line of star-forming galaxies, emitting as much as $0.65\%$ of the typical $L_{\rm FIR} {\sim} 10^{12} \Lsun$ far-IR luminosity of the host \citep{2010ApJ...724..957S}. There is a well-established log-linear relation between \cii\ and the star formation rate (SFR) \citep{2014A&A...570A.121P, 2014A&A...568A..62D, 2015ApJ...800....1H} which becomes more complex at high redshift \citep{2015Natur.522..455C, 2015ApJ...813...36V, 2018A&A...609A.130L}, and as a function of metallicity \citep{2017ApJ...845...96C, 2017ApJ...834....5S}.

The analysis in \citet{2018MNRAS.478.1911P} estimates the \cii\ surface brightness as $I_\nu = 6.6^{+5.0}_{-4.8} \times 10^4$\,Jy/sr, but finds inconclusive support for a \cii\ emission component from a Bayesian model test. This is a tantalizing suggestion of \cii\ at \zrange, and one of our goals here is to attempt to upgrade it to a secure detection or more stringent limit. The BOSS-North region is well out of the galactic plane, but the 545\,GHz map does have significant Milky Way dust emission. This contamination suggests that additional effort toward suppressing the foregrounds may be fruitful. With this as a backdrop, we will describe the general problem of rejecting bright Milky Way continuum foregrounds while constraining correlated line and continuum emission.

Intensity mapping is typically described in terms of one isolated line, such as \cii\ above. However, the target galaxies also contribute a characteristic spectral energy distribution, which includes bright dust or synchrotron continuum contribution at most frequencies of interest for intensity mapping. The overdensity field traced by a companion galaxy redshift survey will, therefore, have non-zero cross-correlation with the intensity survey across many frequencies. The cross-correlation of the intensity survey at $\nu_i$ with galaxies at $z_j$ therefore measures the spectral energy distribution (SED) of galaxies at $z_j$ emitting into $\nu_i$, ${\rm SED}(\nu_i, z_j)$ \citep{2014A&A...570A..98S, 2018MNRAS.478.1911P, 2017ApJ...838...82S, 2018arXiv181000885C}.

A determination of the line brightness must marginalize over the continuum part of the SED. Traditional foreground cleaning that removes uncorrelated continuum emission (such as from the Milky Way) will also suppress the correlated continuum radiation. The cross-correlation in a cleaned map no longer traces ${\rm SED}(\nu_i, z_j)$ but has residual correlated continuum emission due to the slight differences in SED between uncorrelated (largely Milky Way) and correlated continuum radiation. While foreground cleaning may be successful in revealing the redshifted line emission, it may muddle the correlated continuum and produce bias. We argue that intensity mapping analyses should apply a self-consistent approach to deweighting uncorrelated continuum foregrounds and estimating correlated continuum in the target galaxies. 

Estimates of the emission of the Milky Way that are independent of extragalactic sources have utility in removing foregrounds without impacting the extragalactic correlated continuum signal. We find that existing spatial templates of reddening to stars \citep{2015ApJ...810...25G} and inferred from HI \citep{2017ApJ...846...38L} do not significantly aid the line amplitude estimation in the case of Planck 545\,GHz\,$\times$\,BOSS quasars. Similar templates may have great utility in the future for separating continuum terms. 

A fundamental challenge in the Planck analysis is the number of available frequency channels relative to the number of estimated parameters. We consider a future tomographic spectral survey with many frequency bands and argue that low $k_\parallel$ bins isolate the correlated continuum emission (barring instrumental effects). Cutting these bins has a minor impact except in reducing sensitivity at low $|k|$. These conclusions bode well for future spectral intensity mapping surveys \citep{2017arXiv170909066K}, which anticipate numerous frequency channels. Throughout, we will use the example of \cii, but considerations here apply to $21$\,cm and other targets of intensity mapping.

Section\,\ref{sec:lincomb} reviews Milky Way foregrounds in the Planck data, and describes a simple map-space cleaning approach. This approach neglects the joint interpretation of uncorrelated continuum emission from the Milky Way and correlated continuum emission from the target galaxies. Section\,\ref{sec:sensitivity} develops some general results for intensity mapping with both correlated and uncorrelated continuum emission in a simplified setting. It argues for the utility to Milky Way-only templates, which we apply to the Planck cross BOSS analysis in Section\,\ref{sec:mwpxb}. Marginalizing over a correlated continuum model amounts to throwing out those degrees of freedom along the line of sight. An intensity survey should therefore have many more bands than correlated continuum model parameters. Section\,\ref{sec:tomographic} considers this true tomographic case and shows that correlated continuum bias lies at low $k_\parallel$.

\section{Milky Way foregrounds in the Planck data}
\label{sec:lincomb}

\begin{figure}
  \includegraphics[scale=0.45]{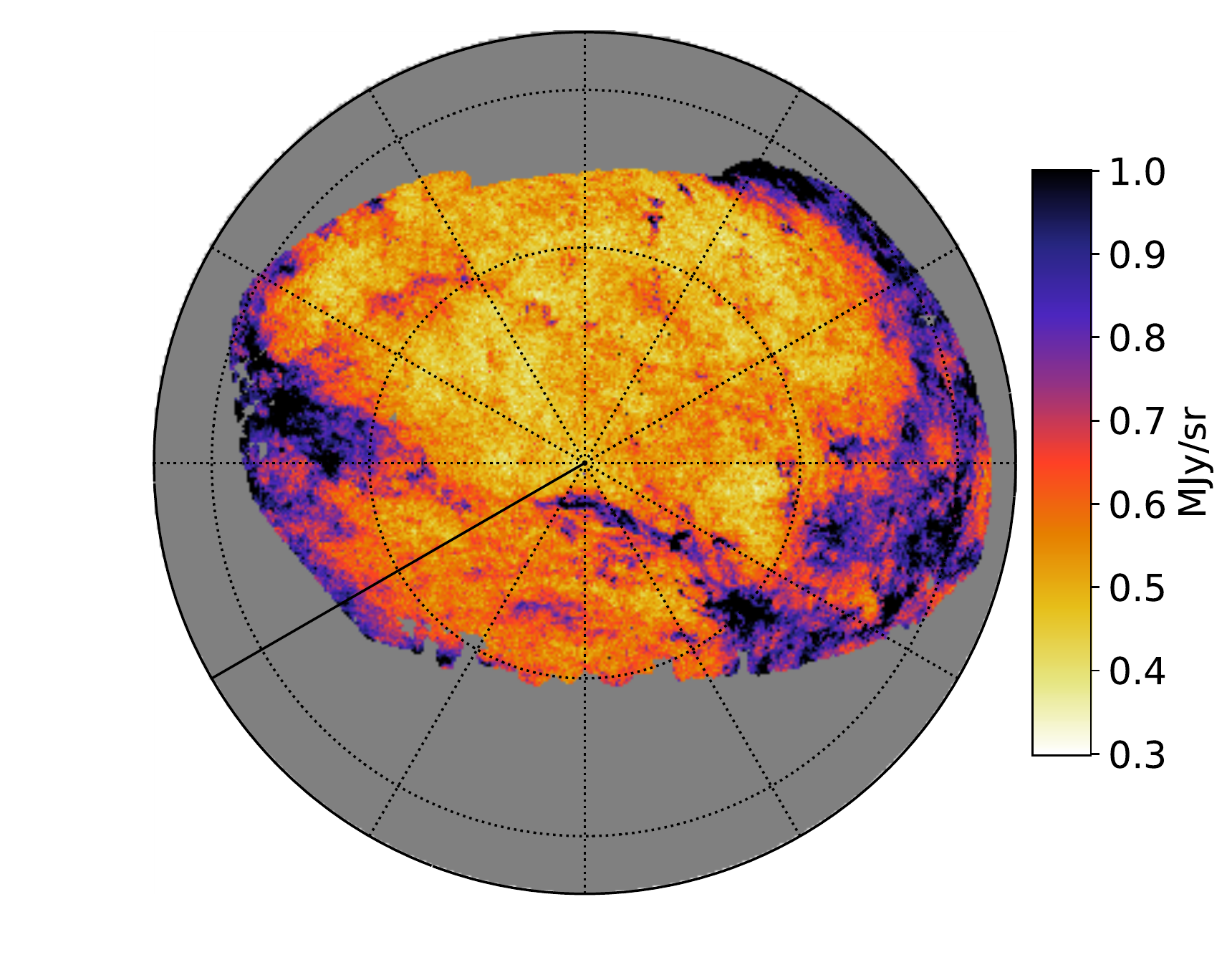}
	\caption{Planck 545\,GHz data in the BOSS-North quasar mask region has significant Milky Way contamination that adds variance to the cross-power with the BOSS quasars. The color bar saturates at $1$\,MJy/sr to show the galactic emission structure.}
    \label{fig:planck545map}
\end{figure}

\begin{figure}
  \includegraphics[scale=0.45]{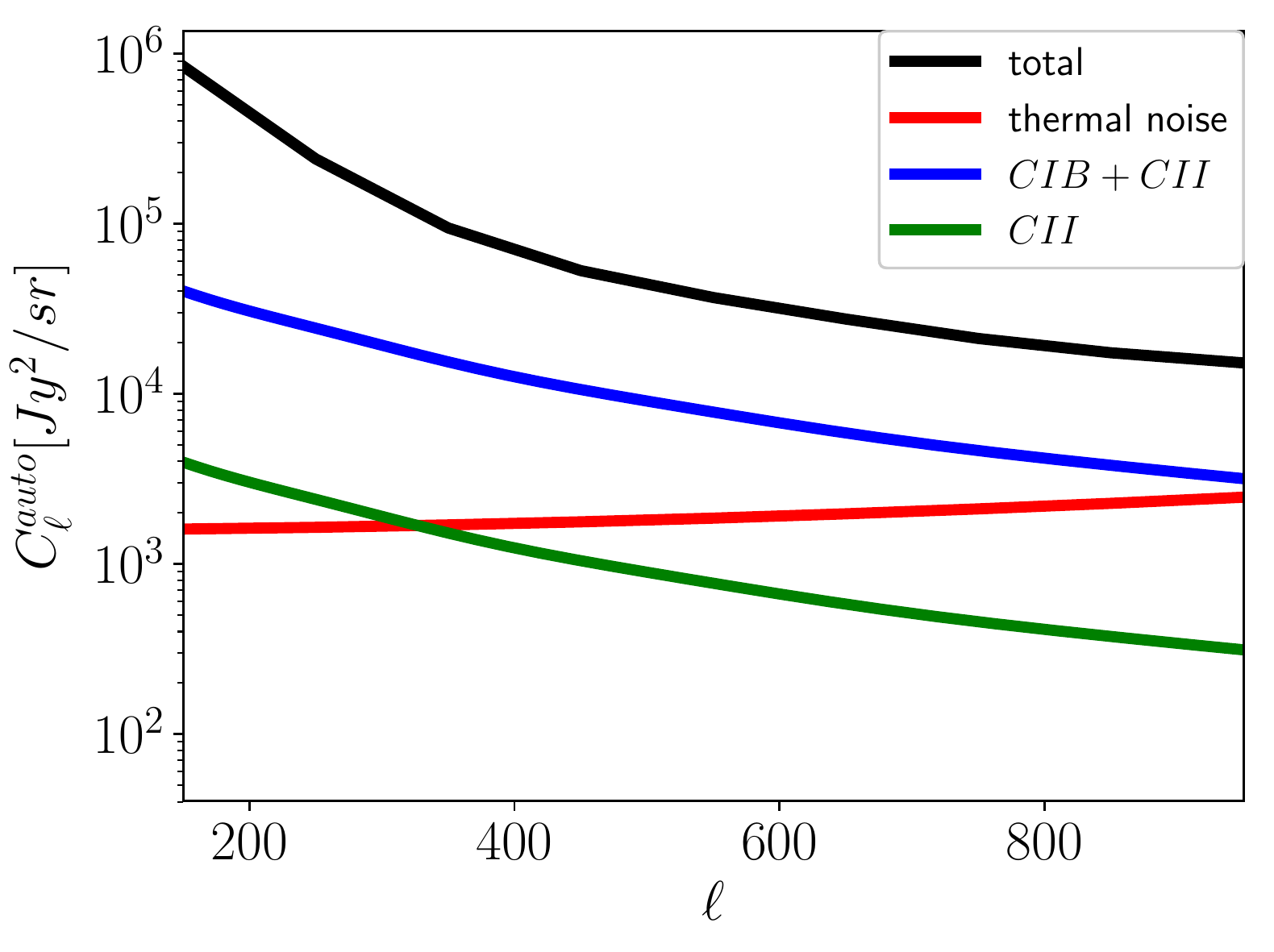}
	\caption{Contributions to the Planck 545\,GHz auto-power in the BOSS survey region. The CIB and \cii\ emission are based on the model and fits in \citet{2018MNRAS.478.1911P} and the ``total" is the measured 545\,GHz auto-power. The line intensity signal is $100\times$ lower than galactic foregrounds and $\approx 10\times$ lower than the correlated ``CIB" continuum. Instrumental noise is lower than foreground power for $\ell <2000$.}
    \label{fig:planck545power}
\end{figure}

\citet{2018MNRAS.478.1911P} present a preliminary indication of \cii\ emission from \zrange\ that is correlated with large-scale structure traced by quasars in the BOSS survey. Planck High Frequency Instrument (HFI) data \citep{2010A&A...520A...9L, 2011A&A...536A...6P} at 545\,GHz serve as the intensity map, and measurements at 353\,GHz and 857\,GHz help to isolate dust continuum and tSZ emission from the \cii\ of interest. Fig.\,\ref{fig:planck545map} shows significant contamination from the Milky Way in the BOSS-North region of the Planck 545\,GHz map. This Milky Way emission is uncorrelated with the \cii\ signal from \zrange, so contributes variance but not bias to the cross-correlation with quasars. Fig.\,\ref{fig:planck545power} shows that the \cii\ signal from \citet{2018MNRAS.478.1911P} remains $100\times$ lower than the total 545\,GHz power spectrum, which is dominated by foregrounds at all scales.

\citet{2018MNRAS.478.1911P} perform no foreground cleaning in the map domain, so the bright Milky Way emission contributes significant variance to each of the cross-powers of 353, 545, and 857\,GHz with BOSS. Our primary goal here is to assess whether some form of foreground cleaning could upgrade the modest indication of \cii\ in \citet{2018MNRAS.478.1911P} to a secure detection of redshifted line emission. We will start with a standard approach of cleaning Milky Way (uncorrelated) continuum emission in map space and argue that this is complicated by correlated dust continuum emission from the target galaxies.

A simple approach to cleaning galactic emission is to form a linear combination of adjacent bands, as 
\begin{equation}
{\bf x}_{\rm clean} = {\bf x}_{545} - \kappa_{353} {\bf x}_{353} - \kappa_{857} {\bf x}_{857}.
\label{eqn:lincleanmap}
\end{equation}

\begin{figure} 
  \includegraphics[scale=0.45]{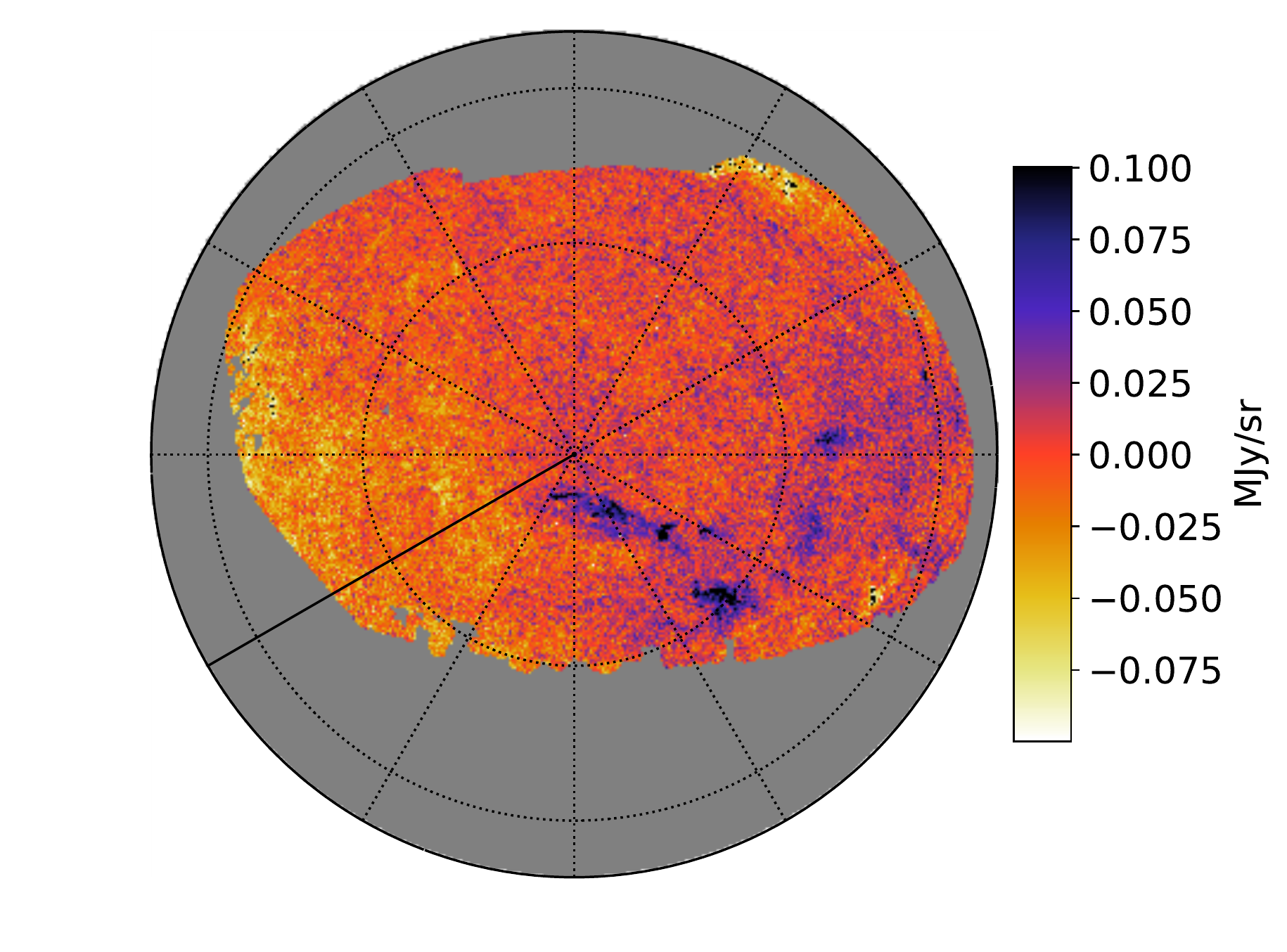}
	\caption{Planck 545\,GHz after cleaning with a linear combination of 353 and 857\,GHz. The colorbar saturates at $0.1$\,MJy/sr to show the residual galactic emission structure, which is greatly reduced from Fig.\,\ref{fig:planck545map}.}
    \label{fig:planck545map_cleaned}
\end{figure}

The goal of finding linear combination parameters $\{ \kappa_{353}, \kappa_{857} \}$ is to minimize the auto-power or RMS variation of the cleaned map (and hence the error bars of the cross-correlation). Stack the $353$\,GHz and $857$\,GHz maps into a matrix ${\bf A}$. Then the linear combination coefficients which minimize variance are given by
\begin{equation}
{\boldsymbol \kappa} = ({\bf A}^T {\bf N}^{-1} {\bf A})^{-1} {\bf A}^T {\bf N}^{-1} {\bf x}_{\rm 545}
\end{equation}
where ${\bf N}$ is the covariance of the $545$\,GHz map. Fig.\,\ref{fig:planck545map_cleaned} shows the map after linear combination cleaning, with some residual Milky Way dust emission still visible, but greatly reduced.

\begin{figure} 
  \includegraphics[scale=0.55]{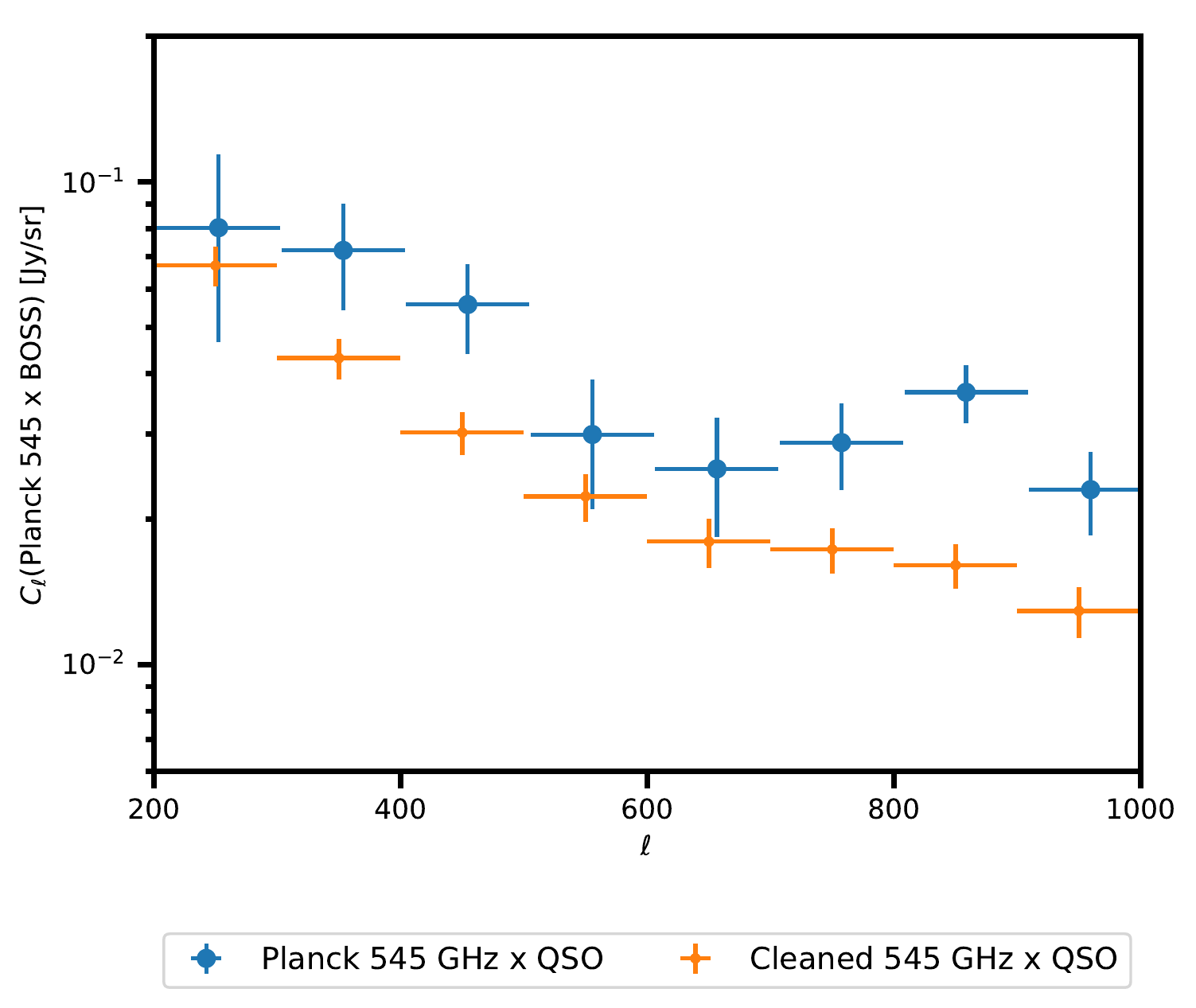}
  \caption{Measured cross-power for Planck $545$\,GHz\,$\times$\,BOSS quasar overdensity in the BOSS-North region, both before and after linear combination cleaning. Error bars after cleaning are a factor of $3$ lower. The linear combination primarily removes uncorrelated Milky Way emission. Much of the correlated continuum emission of the target galaxies is also removed, due to its spectral similarity to the Milky Way dust continuum. Hence in addition to lower errors, the cross-power also shifts down after cleaning.
  \label{fig:cleaned_power}
}
\end{figure}

Fig.\,\ref{fig:cleaned_power} shows the cross-power of quasars with the cleaned map compared to the input 545\,GHz map.\footnote{An internal linear combination can be biased by spurious correlations between signal and foreground, especially on large angular scales. Simulations show $<1\%$ bias for $\ell > 200$ (range reported here), but as we will argue, this map-domain cleaning is primarily illustrative and is not used in parameter estimates.} The dramatic change in the cross-power from linear combination cleaning motivates a reconsideration of foregrounds in the \citet{2018MNRAS.478.1911P} analysis. Not only do the errors diminish, but the level of the cross-power also drops considerably. This is because much of the cross-correlation between Planck 545\,GHz and BOSS quasars is from correlated continuum emission. 

The linear combination that cleans the Milky Way from the 545\,GHz map also removes much of the correlated continuum emission. If the dust continua emission shared the same SED, the linear combination cleaning would be highly effective in removing Milky Way foregrounds, and allow a unique determination of the line brightness independently of correlated continuum emission. The fact that the Milky Way dust SED (which drives the linear combination coefficients) is not identical to the correlated continuum SED means that the cleaned cross-power in Fig.\,\ref{fig:cleaned_power} has some poorly-determined residual correlated continuum emission. 

The inference of clustered \cii\ has two parts 1) a determination of the amplitude of \cii\ plus dust continuum (CIB) clustering, 2) a marginalization over the correlated dust continuum to separate the \cii\ contribution. The clustering of \cii\ emission with quasars in \zrange\ is localized to the 545\,GHz band, so a linear combination of bands that cleans 545\,GHz is an effective way to isolate \cii. In contrast, the marginalization over correlated dust continuum depends on information from 353 and 857\,GHz, so becomes entangled with attempts to use those bands to clean Milky Way emission at 545\,GHz.

\section{Sensitivity with continuum emission}
\label{sec:sensitivity}

This section builds up a set of analytic results for multi-tracer analysis of intensity mapping data with correlated and uncorrelated continuum emission contamination. It has become standard to clean continuum contamination in  intensity mapping data in the map domain, and to simulate the impact of resulting signal loss on the power spectrum (e.g. \citet{2015ApJ...815...51S} for single dish instruments and \citet{2018ApJ...868...26C} for interferometers). In contrast, the multi-tracer approach developed here jointly 1) deweights Milky Way foregrounds, 2) marginalizes over correlated continuum emission, and 3) estimates the line amplitude. In this context, signal loss does not need to be estimated separately because it is included self-consistently in the parameter estimate. Rather than calculating signal loss through simulations, expressions here will focus on the error of the line estimate, and how continuum emission causes constraints to degrade from thermal noise limits.

For the expressions to remain tractable and illustrative, we will make several simplifying assumptions: 1) shot noise in the galaxy redshift survey is negligible and the linear bias is known, 2) inter-bandpower correlations are negligible so we can consider expressions for a single bandpower/multipole, 3) there is no stochasticity between the density traced in the intensity and galaxy redshift survey. The line intensity and galaxy fields should trace the same underlying density on linear scales, making this approximation reasonable on large-area surveys such as used in Section\,\ref{sec:mwpxb}. 

This regime is applicable to single-dish surveys that can access linear modes, such as BINGO \citep{2016arXiv161006826B}, GBT \citep{2013MNRAS.434L..46S}, Parkes \citep{2018MNRAS.476.3382A}, and MeerKAT/SKA \citep{2015ApJ...803...21B} for 21\,cm, and most of the proposed experiments for high-frequency lines: CCAT-prime \citep{2018SPIE10708E..1UV}, COMAP \citep{2018arXiv180807487T}, CONCERTO \citep{2018arXiv180204804D}, EXCLAIM\footnote{Eric Switzer, private communication}, SPHEREx \citep{2014arXiv1412.4872D}, STARFIRE \citep{2014ApJ...793..116U}, and HETDEX \citep{2017MNRAS.464.1948F}. Some of these simplifying assumptions can be easily relaxed in this framework. Appendix\,\ref{apx:multi-tracer} describes the impact of stochasticity and shot noise, and Section\,\ref{ssec:loscorr} describes low $k_\parallel$ modes. Additionally, the spherical harmonic approach here is well-matched to surveys on large (curved) regions of the sky.

\subsection{Uncorrelated foregrounds in a multi-tracer setting}

Foregrounds that are uncorrelated with cosmological density will increase the errors of a determination of the line brightness from cross-correlation, but not bias the result. This section quantifies the impact of uncorrelated foregrounds in a multi-tracer \citep{2009JCAP...10..007M} setting, reviewed in Appendix\,\ref{apx:multi-tracer}.

Let ${\boldsymbol \delta}$ (a vector) be a spatial map of the cosmological overdensity in a redshift slice that corresponds to a frequency channel in the intensity survey. Model the observed galaxy redshift survey map ${\bf x}_g$ as a tracer with linear bias $b_g$ and shot noise ${\bf n}_{g}$. Model the intensity map as a line surface brightness amplitude $S_L$ times overdensity ${\boldsymbol \delta}$ plus noise ${\bf n}_I$. (Here $S_L = I_L b_L$ refers to the line brightness $I_L$ multiplied by the linear cosmological bias $b_L$ of the line emission.) Emission from the Milky Way and galaxies not in the selected redshift slice will contribute continuum emission that is uncorrelated with ${\boldsymbol \delta}$. Let the uncorrelated continuum foregrounds have a spatial template ${\bf x}_{u}$ and spectral dependence $f(\nu)$. These appear in the intensity map at frequency $\nu_I$ as $f(\nu_I) {\bf x}_{u}$. All vectors refer to 2D map slices at constant frequency or redshift. The observed galaxy redshift survey (${\bf x}_g$) and intensity (${\bf x}_I$) maps are
\begin{eqnarray}
{\bf x}_{g} &=& b_g {\boldsymbol \delta} + {\bf n}_g \nonumber \\
{\bf x}_{I} &=& S_L {\boldsymbol \delta} + f(\nu_I) {\bf x}_{u} + {\bf n}_I.
\label{eqn:uncorrfgonly}
\end{eqnarray}

Form a data vector ${\bf d} = \{ C^{gI}_\ell, C^{gg}_\ell \}$ from the power spectrum of the galaxy redshift survey $C^{gg}_\ell$ and the cross-power between the intensity and galaxy redshift survey maps $C^{gI}_\ell$. 
Model the observed two-point functions as
\begin{eqnarray}
C^{gI}_\ell &=& b_g S_L C^{\delta \delta}_\ell \nonumber \\
C^{gg}_\ell &=& b_g^2 C^{\delta \delta}_\ell + N_g.
\end{eqnarray}
where the galaxy bias and shot noise are assumed to be known, for simplicity. The model then has parameters ${\boldsymbol \theta} = \{ S_L, C^{\delta \delta}_\ell \}$ where $C^{\delta \delta}_\ell$ is the power spectrum of underlying density ${\boldsymbol \delta}$ fluctuations.
For illustrative purposes we consider the constraint based on a single multipole or bandpower. The log-likelihood is
\begin{equation}
2 \mathcal{L} = \ln \det {\boldsymbol \Sigma}({\boldsymbol \theta}) + [{\bf d} - {\boldsymbol \mu}({\boldsymbol \theta})]^T {\boldsymbol \Sigma}^{-1}({\boldsymbol \theta}) [{\bf d} - {\boldsymbol \mu}({\boldsymbol \theta})].
\label{eqn:likelihood}
\end{equation}

The terms in the covariance ${\boldsymbol \Sigma}$ under the assumption of Gaussian fluctuations in the data are given by 
\begin{equation}
{\rm Cov}(C^{AB}_\ell, C^{CD}_{\ell'}) = \frac{1}{v_\ell} \delta_{\ell, \ell'} ( C^{AD}_\ell C^{BC}_\ell + C^{AC}_\ell C^{BD}_\ell ),
\label{eqn:simplecov}
\end{equation}
where $C^{AB}_\ell$ is the cross-power of fields $A$ and $B$. $v_\ell$ is the number of modes in the measurement, which is roughly $(2 \ell + 1) f_{\rm sky} \Delta \ell$ for a 2D map on the fraction $f_{\rm sky}$ of the sky and a bandpower of width $\Delta \ell$. The delta function $\delta_{\ell, \ell'}$ is exact for all-sky, and for simplicity we assume it is approximately the case for broad bandpowers on a partial sky.

Following the approach of Appendix\,\ref{apx:multi-tracer}, the variance of a determination of the line amplitude $S_L$ from a single multipole $\ell$ is (marginalized over $C^{\delta \delta}_\ell$)
\begin{equation}
\sigma^2_{S_L} = \frac{N_I + f(\nu_I)^2 C^{uu}_\ell}{C^{\delta \delta}_\ell v_\ell},
\label{eqn:fishfg}
\end{equation}
where $C^{uu}_\ell$ is the power spectrum of the uncorrelated emission (the map ${\bf x}_u$), $N_I$ is the noise variance of the intensity map, and $C^{\delta \delta}_\ell$ is the power spectrum of the density fluctuations ${\boldsymbol \delta}$. This expression is exact (neglecting shot noise), rather than an expansion about small intensity map noise \citep{2011MNRAS.416.3009B}. From Eq.\,\ref{eqn:fishfg}, foregrounds that are uncorrelated with the cosmological signal play the same role as noise in the intensity map. 

Cosmic variance would appear in $\sigma^2_{S_L}$ as a term $\propto S_L^2$. As described in Appendix\,\ref{apx:multi-tracer}, this is avoided by adding covariance to the galaxy auto-power $C^{gg}_\ell$ in the likelihood (assuming negligible stochasticity between the intensity and galaxy redshift survey). 

\subsection{Deweighting uncorrelated continuum emission}
\label{ssec:fisherwithfg2band}

Intensity measurements at two frequencies can separate uncorrelated continuum and line emission. Include a ``veto'' map ${\bf x}_V$ at a different frequency $\nu_V$. The uncorrelated continuum foreground $f(\nu_I) {\bf x}_u$ in the intensity map will appear at this frequency as $f(\nu_V) {\bf x}_{u'}$, where $f(\nu_V)$ accounts for the SED and $u'$ in ${\bf x}_{u'}$ denotes that the spatial pattern of uncorrelated foregrounds in the veto map may not be fully coherent with the spatial pattern ${\bf x}_{u}$ in the intensity map. Model this set of maps as
\begin{eqnarray}
{\bf x}_{g} &=& b_g {\boldsymbol \delta} + {\bf n}_g \nonumber \\
{\bf x}_{I} &=& f(\nu_I) {\bf x}_{u} + S_L {\boldsymbol \delta} + {\bf n}_I \nonumber \\
{\bf x}_{V} &=& f(\nu_V) {\bf x}_{u'} + {\bf n}_V.
\label{eqn:mapsfgonly}
\end{eqnarray}
The veto map will also contain line signal from a different redshift, given as $S_L(z_V) {\boldsymbol \delta}(z_V)$, but we will assume that this has negligible correlation with ${\boldsymbol \delta}(z_I)$ in the primary intensity survey. That is an excellent approximation for e.g. \cii\ emission in Planck $353$\,GHz versus $545$\,GHz, which originate from considerably different redshifts. In this case, the redshifted line emission in the veto band amounts to uncorrelated noise, so is accommodated in the assumed noise power $N_V$. Two bands are sufficient to fit for a line amplitude and one continuum parameter. 

Intensity surveys with greater adjacency in the bands must account for signal correlations in the likelihood. Standard codes \citep{2013JCAP...11..044D, 2011PhRvD..84d3516C} can calculate inter-band signal correlations, and several analyses of large-scale structure \citep{2012MNRAS.422.2904G, 2014PhRvD..90f3515N, 2014JCAP...01..042D, 2017MNRAS.468.2938S} have used similar approaches that could be adopted for intensity mapping. Appendices \ref{app:sepfg} and \ref{app:multiband} describe the multi-band case, which can accommodate more general spectral energy distributions of the continuum correlations and inter-band signal correlations. Section\,\ref{ssec:loscorr} and Section\,\ref{sec:tomographic} describe low $k_\parallel$ modes in the $C_\ell$ and $P(k_\parallel, k_\perp)$ context, respectively.

To characterize the lack of coherence between the spatial distribution of uncorrelated foreground contamination in the intensity map (${\bf x}_u$) and in the veto map (${\bf x}_{u'}$), define ``foreground" stochasticity $r_F = C^{uu'}_\ell / \sqrt{C^{uu}_\ell C^{u' u'}_\ell}$, where $C^{uu'}_\ell$ is the cross-power between ${\bf x}_{u}$ and ${\bf x}_{u'}$. 

Expand the observation vector in Eq.\,\ref{eqn:likelihood} to include the cross-power of the veto map and galaxy redshift survey ${\bf d} = \{ C^{gI}_\ell, C^{gV}_\ell, C^{gg}_\ell \}$, and estimate the parameters ${\boldsymbol \theta} = \{ S_L, C^{\delta \delta}_\ell \}$. In the limit that the veto band measures the foregrounds well, e.g. $N_V \ll f(\nu_V)^2 C^{uu}_\ell$. Then,
\begin{eqnarray}
\sigma^2_{S_L} &=& \frac{1}{v_\ell C^{\delta \delta}_\ell} \left [ N_I + \frac{f(\nu_I)^2}{f(\nu_V)^2} N_V \right ] \nonumber \\
&&+ \frac{f(\nu_I)^2 C^{uu}_\ell}{v_\ell C^{\delta \delta}_\ell}(1 - r_F^2).
\label{eqn:cleanedsl}
\end{eqnarray}

To interpret this result, make a cleaned map that removes uncorrelated continuum emission using the veto map, as ${\bf x}_{\rm clean} = {\bf x}_{I} - [f(\nu_I) / f(\nu_V)] {\bf x}_{V}$. In the limit that the foreground in the veto band perfectly traces the intensity map ($r_F=1$), then ${\bf x}_{\rm clean} = S_L {\boldsymbol \delta} + {\bf n}_I - [f(\nu_I) / f(\nu_V)] {\bf n}_V$. The noise variance in this cleaned map is $N_I + N_V [f(\nu_I)^2 / f(\nu_V)^2]$, which is the numerator of the first term in Eq.\,\ref{eqn:cleanedsl}. The first term therefore describes the impact of thermal noise in a cleaned map. To interpret the second term, note that the numerator is $f(\nu_I)^2 C^{uu}_\ell$, the foreground covariance in the intensity map. If $r_F=0$, the veto band provides no cleaning in the intensity band, and the variance is increased by the full foreground brightness analogously to Eq.\,\ref{eqn:fishfg}. Alternately, if $r_F=1$, the veto band is a perfect tracer of uncorrelated foregrounds in the intensity map and the second term goes to zero. In this limit, the line brightness can be determined without impact from the foreground variance.

We have not explicitly cleaned the map anywhere in the likelihood. The appearance of an underlying cleaned map ${\bf x}_{\rm clean}$ arises self-consistently with the parameter estimate through the action ${\boldsymbol \Sigma}^{-1} {\bf d}$ in the likelihood, Eq.\,\ref{eqn:likelihood}. Foregrounds common to both the intensity and veto map show up as correlated noise between the crosspowers $C^{gI}_\ell$ and $C^{gV}_\ell$. Specifically, these off-diagonal terms in ${\boldsymbol \Sigma}$ appear from the cross-variance $C^{IV}_\ell = f(\nu_I) f(\nu_V) C^{uu'}_\ell$. For a simple demonstration in the limit that $r_F=1$ and there is no noise in the veto map ($N_V = 0$), ${\boldsymbol \Sigma}^{-1}$ takes a linear combination of the cross-powers $C^{gI}_\ell$ and $C^{gV}_\ell$ to form a cleaned cross-power analogous to the cleaned map, as
\begin{equation}
C^{gI}_\ell |_{\rm clean} = C^{gI}_\ell - [f(\nu_I) / f(\nu_V)] C^{gV}_\ell. 
\end{equation}
An additional map cleaned through a linear combination in map space will not add information to the likelihood. Furthermore, taking a linear combination such as Eq.\,\ref{eqn:lincleanmap} outside of the context of the likelihood can induce bias from e.g. spurious correlation of foregrounds and signal. This necessitates signal loss simulations to account for the impact of any map operations done before the parameter likelihood in power-spectrum space \citep{2015ApJ...815...51S}. 

In summary, bright Milky Way continuum foregrounds can be deweighted through observations at multiple frequencies, limited by the coherence $r_F$ between frequencies.

\subsection{Uncorrelated and correlated continuum emission}
\label{ssec:corrcontimpact}

In addition to line emission, the host galaxies also emit a dust continuum that correlates with cosmological overdensity. To determine the line amplitude independently of the continuum in these galaxies, the continuum contribution must be modeled and marginalized over. Model this correlated continuum component as $g(\nu) S_C {\boldsymbol \delta}$, where $g(\nu)$ describes the SED of the correlated continuum and $S_C$ is the amplitude of the continuum. Extending the map model in Eq.\,\ref{eqn:mapsfgonly},
\begin{eqnarray}
{\bf x}_{g} &=& b_g {\boldsymbol \delta} + {\bf n}_g \nonumber \\
{\bf x}_{I} &=& f(\nu_I) {\bf x}_{u} + g(\nu_I) S_C {\boldsymbol \delta} + S_L {\boldsymbol \delta} + {\bf n}_I \nonumber \\
{\bf x}_{V} &=& f(\nu_V) {\bf x}_{u'} + g(\nu_V) S_C {\boldsymbol \delta} + {\bf n}_V.
\label{eqn:totalmodel}
\end{eqnarray}

Continue to use the observation vector ${\bf d} = \{ C^{gI}_\ell, C^{gV}_\ell, C^{gg}_\ell \}$, but expand the parameters to include an estimate of the correlated continuum amplitude $S_C$, as ${\boldsymbol \theta} = \{ S_L, S_C, C^{\delta \delta}_\ell \}$. The variance of the estimate of $S_L$, not marginalized over the other parameters ($(F_{S_L, S_L})^{-1}$ for Fisher matrix $F$) is the same as Eq.\,\ref{eqn:cleanedsl}. That is, we can determine the amplitude of ${\boldsymbol \delta}$ in the intensity map as before. However, now the correlated amplitude is $S_L + g(\nu_I) S_C$. To determine the line brightness $S_L$ independently of the correlated continuum $g(\nu_I) S_C$, we need to marginalize over $S_C$, giving 
\begin{eqnarray}
\sigma^2_{S_L} &=& \frac{1}{v_\ell C^{\delta \delta}_\ell} \left [N_I + \frac{g(\nu_I)^2}{g(\nu_V)^2} N_V \right ] \nonumber \\
&&+ \frac{C^{uu}_\ell}{v_\ell C^{\delta \delta}_\ell}  \left [ f(\nu_I) - f(\nu_V) \frac{g(\nu_I)}{g(\nu_V)} \right ]^2, \nonumber \\
&&+ \frac{2 C^{uu}_\ell}{v_\ell C^{\delta \delta}_\ell} (1-r_F) f(\nu_I) f(\nu_V ) \frac{g(\nu_I)}{g(\nu_V)}. 
\label{eqn:totalerr}
\end{eqnarray}

To interpret this result, make a cleaned map that removes correlated continuum emission using the veto map, as ${\bf x}_{\rm clean} = {\bf x}_{I} - [g(\nu_I) / g(\nu_V)] {\bf x}_{V}$. Unlike the previous section, this map cleans the correlated continuum emission using a linear combination of bands. The first term in Eq.\,\ref{eqn:totalerr} is just the thermal noise in this cleaned map. The term in brackets in the second line is the residual of the uncorrelated continuum foreground emission after the correlated emission has been cleaned. If both the correlated and uncorrelated continuum have the same SED, $g(\nu) \propto f(\nu)$, both are deweighted equally. The third term is a cross-term, and arises from the stochasticity between the spatial shape of uncorrelated continuum foregrounds in maps ${\bf x}_I$ and ${\bf x}_V$, which contributes variance to $\sigma^2_{S_L}$ even if the correlated and uncorrelated continuum have the same SED. 

\subsection{Bias from incorrect correlated continuum models}

If the correlated continuum spectrum is not accurately modeled, residuals may be spuriously interpreted as line emission. In the linear model for two intensity bands (Eq.\,\ref{eqn:totalmodel}), let the true correlated continuum SED be $g'(\nu)$ and model it as $g(\nu)$. In this case, the estimated line amplitude is 
\begin{equation}
\hat S_L = S_L + S_C \left [ g'(\nu_I) - \frac{g(\nu_I)}{g(\nu_V)} g'(\nu_V) \right ],
\end{equation}
which is biased by the difference between the true SED and the model. 

In the case where the number of instrument bands $N_{\rm band}$ equals the number of parameters $N_{\rm param}$, the spectral model uses all degrees of freedom and there is no additional handle to identify bias in the line amplitude. The model may have a excellent goodness of fit because it is using $S_L$ to fit a residual from an insufficient continuum model. If $N_{\rm band} \gg N_{\rm param}$, channels that have no expected line emission correlation should be consistent with zero, providing a test.

In summary, in the case of relatively few bands, the line amplitude may be undetectably compromised by an incomplete continuum SED model, and subject to variance from bright Milky Way emission which cannot be independently downweighted. Section\,\ref{sec:manybands} develops a model to forecast an experiment with many bands in a $C_\ell$ approach. Section\,\ref{sec:tomographic} describes correlated continuum emission in the context of $P(k_\perp, k_\parallel)$ in a survey with many bands.

\section{Extension to many bands}
\label{sec:manybands}

The preceding toy models give some intuition for the impact of uncorrelated and correlated continuum emission. The analytic Fisher matrix approach presented there becomes cumbersome to treat more than two intensity channels, requiring analytic $N_{\rm band} \times N_{\rm band}$ inverses for $N_{\rm band}$ maps for each parameter, followed by an analytic $N_{\rm param} \times N_{\rm param}$ inverse over parameters to find the marginalized error on $S_L$. 

Appendix\,\ref{app:multiband} builds a linear model for the joint deweighting of uncorrelated continuum and fit to both correlated line and continuum emission. Here we model the cross-power at each $\ell$ as a set of correlated continuum and line parameters ${\boldsymbol \theta}$ which linearly describe the measured cross-powers as a function of frequency ${\bf d}_\ell$ as ${\bf d}_\ell = {\bf M} {\boldsymbol \theta} + {\bf n}_\times$, where the bandpowers have noise ${\bf n}_\times$ described by covariance ${\bf N}_\times$. The parameters ${\boldsymbol \theta} = \{ S_L, {\bf S}_C \}$, where $S_L$ is the line amplitude and ${\bf S}_C|_j = S_{C,j}$ is the amplitude of linear spectral templates $g_j(\nu_i)$ that describe the correlated continuum emission as a function of frequency $\nu_i$. The matrix ${\bf M} \equiv [{\boldsymbol \xi}, {\bf g}_1, ... {\bf g}_{N_{\rm comp}}]$ holds the spectral templates for the line correlation ${\boldsymbol \xi}$ and $N_{\rm comp}$ component spectral modes of the correlated continuum. ${\boldsymbol \xi}_i = \xi(\nu_i)$ describes the correlation of the line intensity at each frequency $\nu_i$ with the galaxies at redshift $z$. In the case of Planck, the bands are widely-spaced in redshift so that $\xi(\nu_i)$ is a delta function at $\nu_i = 545$\,GHz to an excellent approximation. Section\,\ref{ssec:loscorr} describes the impact of correlations at low $k_\parallel$ on an intensity survey with many narrow bands. Appendix\,\ref{app:sepfg} describes how the Milky Way produces strong, low-rank correlations in ${\bf N}_\times^{-1}$, and Appendix\,\ref{app:lowrankgeo} describes the case of foregrounds that have the same spatial template at all frequencies.

The variance ${\boldsymbol \Sigma}_{\boldsymbol \theta}$ on the correlated line and continuum amplitude parameters ${\boldsymbol \theta}$ in this simple linear model is
\begin{equation}
{\boldsymbol \Sigma}_{\boldsymbol \theta} = ({\bf M}^T {\bf N}_\times^{-1} {\bf M})^{-1}.
\label{eqn:multierr}
\end{equation}
Appendix\,\ref{app:multiband} shows that this simple form reproduces the results from previous sections under a slightly more restrictive set of assumptions.

\subsection{How many bands are needed?}
\label{ssec:howmany}

\begin{figure}
  \includegraphics[scale=0.57]{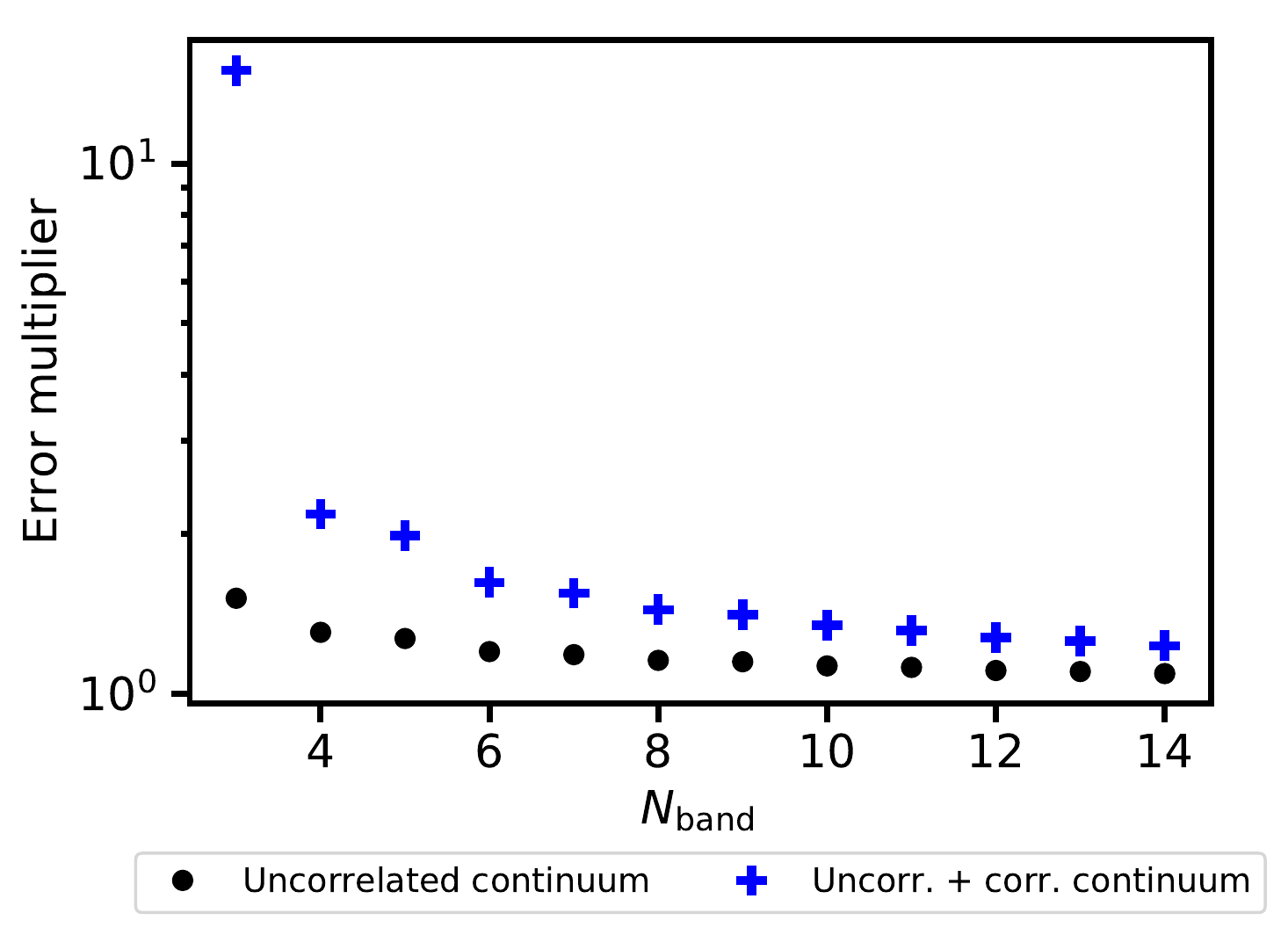}
  \caption{Errors on the determination of line amplitude (divided by thermal noise limits) as a function of the number of bands in a survey with three correlated degrees of freedom (line amplitude, and correlated continuum amplitude and effective temperature). A survey with only three bands has insufficient freedom to suppress Milky Way emission while also marginalizing over the correlated continuum.
  \label{fig:errors_nmode}}
\end{figure}

The linear formulation in Eq.\,\ref{eqn:multierr} emphasizes an accounting for overall degrees of freedom (DOF). Let the line signal have one degree of freedom (the line amplitude $S_L$), the correlated continuum model have $D_{\rm corr}$ DOF, and the uncorrelated continuum emission model have $D_{\rm uncorr}$ DOF. Assuming these modes are mutually independent, the total number of degrees of freedom needed to account for all terms is $D_{\rm corr} + D_{\rm uncorr} + 1$. If a survey has $D_{\rm corr} + 1$ bands, it can separate correlated line and continuum emission, but it has no degrees of freedom remaining to suppress uncorrelated (Milky Way) emission. In the model of Eq.\,\ref{eqn:totalmodel}, there are two intensity map channels, and one degree of freedom for line and correlated continuum emission respectively. This leaves no additional freedom to remove uncorrelated foregrounds, and these show up in the second term of Eq.\,\ref{eqn:totalerr}.

Equation\,\ref{eqn:multierr} permits a rapid simulation of an instrument with $N_{\rm band}$ uniform bands for a hypothetical \cii\ survey from $2 < z < 2.5$ ($543$\,GHz and $633$\,GHz) to illustrate degrees of freedom. For correlated emission, we fit an unknown amplitude for a $T_{\rm dust} = 27.2$\,K, $\beta=1.5$ dust source at $z=2.25$ ($I^{\rm dust}_\nu = \nu^\beta B[T_{\rm dust}, (1+z) \nu]$ where $B(T_{\rm dust}, \nu)$ is the Planck law) and for the amplitude of the derivative of intensity with respect to temperature $dI_\nu^{\rm dust}/dT$, corresponding to a fit to the dust temperature to first order. There are two correlated degrees of freedom to fit the continuum and one signal DOF, which gives three DOF in total. For Milky Way dust emission, we take emission at 19\,K and $\beta=1.5$ with an amplitude that is $10\times$ thermal noise (after marginalizing over the correlated continuum). We take the signal correlation to exist only in one spectral band, but Sec.\,\ref{ssec:loscorr} describes the impact of signal correlations along the line of sight. Figure\,\ref{fig:errors_nmode} shows the inflation of errors (relative to thermal) as a function of the number of bands. In a survey with three bands, there is no remaining latitude to deweight uncorrelated continuum, and the error on the line amplitude is penalized by $10\times$ thermal, consistent with the variance of the contamination. Adding one additional channel provides a degree of freedom to suppress the Milky Way, independently of the correlated continuum determination.

The simulation in this section is set up to emphasize the impact of DOF in a general survey. Sec.\,\ref{ssec:interpmcmc} uses Eq.\,\ref{eqn:multierr} to interpret the Planck cross BOSS and shows that there is not a strong penalty for the three bands in this particular case, given the measured amplitude of the Milky Way foreground, and similarity in the correlated and uncorrelated spectral indices.

\subsection{Line-of sight correlations}
\label{ssec:loscorr}

\begin{figure}
  \includegraphics[scale=0.57]{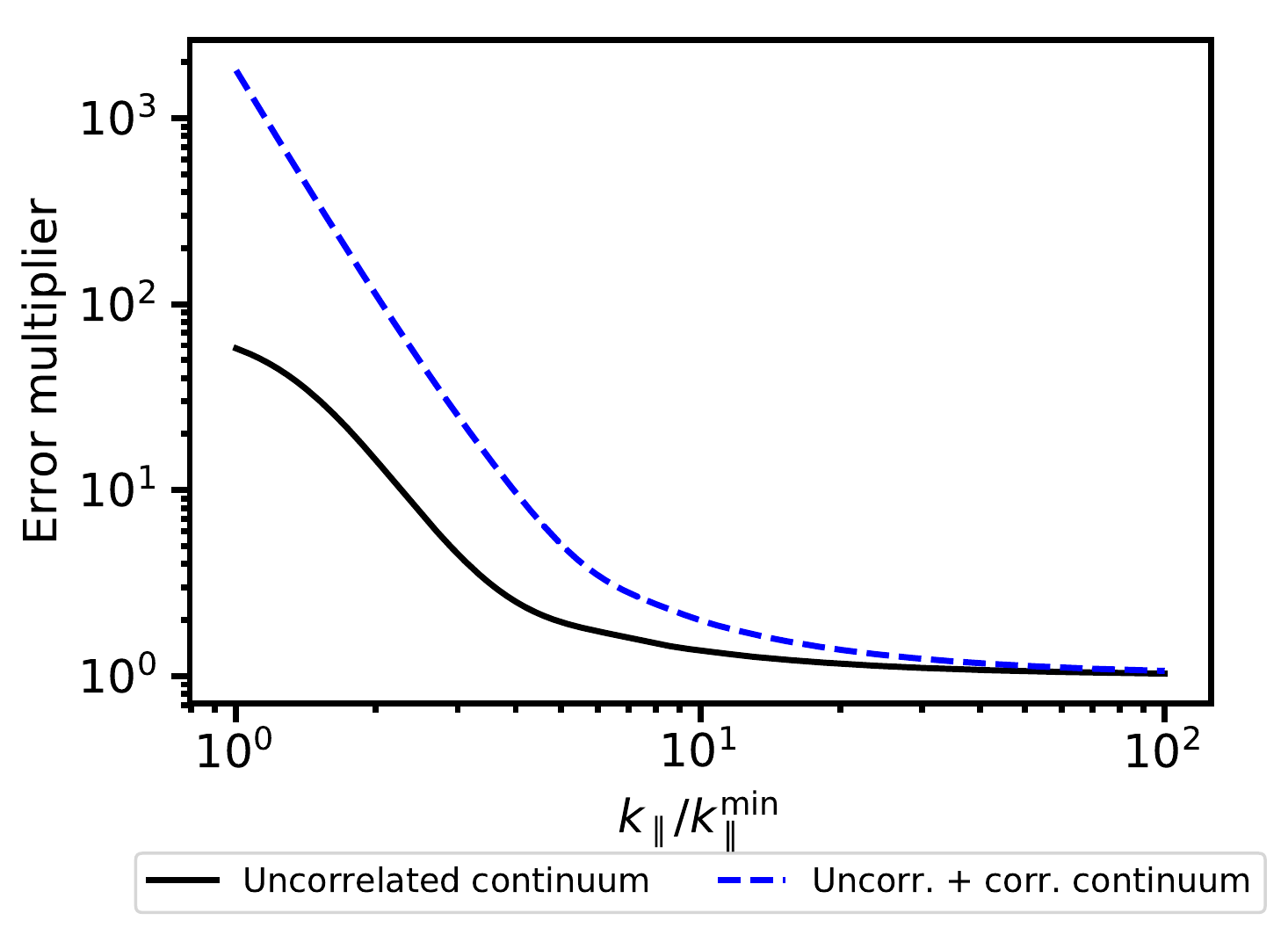}
  \caption{Errors on the determination of line amplitude (divided by thermal noise limits) as a function of the signal  correlation length along the line of sight, expressed in terms of wavenumber relative to the largest accessible in the survey $k_\parallel^{\rm min}$. The lowest $k_\parallel$-modes have smooth spectra that overlap with correlated (extragalactic) and uncorrelated (Milky Way) continuum emission.
  \label{fig:errors_los}}
\end{figure}

Cosmological structure is also correlated along the line of sight, so a slice of density from the galaxy redshift survey at redshift $z$ will correlate not only with line emission from redshift $z$ but also nearby redshifts $z'$. Correlations at low $k_\parallel$ (consequently at low-$\ell$) can extend over many spectral bins in the intensity map. In the case of Planck, this effect is negligible between the bands $353$\,GHz, $545$\,GHz, and $857$\,GHz (at the $\ell$ reported here), but the effect will be significant for future tomographic surveys with many narrow bands.

Sec.\,\ref{sec:tomographic} develops a complete view of this effect by considering the 3D power spectrum of a tomographic survey. It shows that signal at low $k_\parallel$ is corrupted by correlated continuum emission. However, the linear framework in Eq.\,\ref{eqn:multierr} provides another approach to understand the impact of this effect. Appendix\,\ref{app:multiband} describes line-of-sight signal correlations in terms of a correlation kernel $\xi(\nu_i)$ between the line intensity at $\nu_i$ and a galaxy survey at fixed redshift $z$. This kernel can be calculated as described in \citet{2013JCAP...11..044D, 2011PhRvD..84d3516C}.

Fig.\,\ref{fig:errors_los} demonstrates the inflation of errors at low $k_\parallel$, in the same simulation setup as Sec.\,\ref{ssec:howmany} and taking 200 channels. At low $k_\parallel$, the clustered line emission is spectrally smooth. It is, therefore, less distinguishable from correlated and uncorrelated continuum emission, so its errors rapidly grow. Here, modes with spatial wavenumbers $k_\parallel$ $10\times$ higher than the largest modes in the volume gain immunity by varying rapidly spectrally compared to continuum.

It has been recognized since \citet{1979MNRAS.188..791H} that bright and spectrally smooth Milky Way emission contaminates long $k_\parallel$ modes. A new point emphasized here is that correlated continuum emission also contaminates these modes.

\subsection{Application to auto-power analysis}

The formalism and discussion here have applied to the cross-power between an intensity survey and galaxy redshift survey tracer. It is useful to contrast this to considerations for the intensity survey auto-power. In the cross-power, the uncorrelated continuum (Milky Way) adds variance, so the aim is simply to reduce the Milky Way variance. In the auto-power, residual Milky Way variance translates directly into a bias that is difficult to model because the Milky Way is 1) much brighter than the correlated continuum (and so instrumental response is even more critical), 2) is not statistically isotropic, and 3) does not have a uniform spectrum. Our discussion is therefore limited to the cross-power, with possible application to the auto-power deferred to future work.

\subsection{Adding a template map for the Milky Way}
\label{sec:analyticmwtemplate}

The measurement of Planck $\times$ BOSS quasars is in the challenging regime where $N_{\rm band} \approx N_{\rm param}$. Given that this is the best intensity map for \cii\ currently available, we can try to improve the foreground cleaning with a Milky Way template. 

Extend the map model in Eq.\,\ref{eqn:totalmodel} to include a Milky Way template with amplitude $A_M$ and noise variance $N_M$, as
\begin{eqnarray}
{\bf x}_{g} &=& b_g {\boldsymbol \delta} + {\bf n}_g \nonumber \\
{\bf x}_{I} &=& f(\nu_I) {\bf x}_u + g(\nu_I) S_C {\boldsymbol \delta} + S_L {\boldsymbol \delta} + {\bf n}_I \nonumber \\
{\bf x}_{V} &=& f(\nu_V) {\bf x}_u + g(\nu_V) S_C {\boldsymbol \delta} + {\bf n}_V \nonumber \\
{\bf x}_{M} &=& A_M {\bf x}_u + {\bf n}_M.
\label{eqn:totalmodelwithtemplate}
\end{eqnarray}

Eq.\,\ref{eqn:multierr} gives 
\begin{eqnarray}
\sigma^2_{S_L} &=& \frac{1}{v_\ell C^{\delta \delta}_\ell} \left [ N_I + \left [ \frac{g(\nu_I)}{g(\nu_V)} \right ]^2 N_V \right ] \nonumber \\
&&+ \frac{N_M A_M^{-2}}{ v_\ell C^{\delta \delta}_\ell} \left [f(\nu_I) - f(\nu_V) \frac{g(\nu_I)}{g(\nu_V)} \right ]^2.
\label{eqn:totalwtemplate}
\end{eqnarray}
This is just Eq.\,\ref{eqn:totalerr} where the bright Milky Way variance $C^{uu}_\ell$ has been traded with noise-to-signal ratio in the tracer map, $N_M /A_M^2$. Appendix\,\ref{sec:fishermw} includes the impact of stochasticity $r_M$ between the tracer of the Milky Way emission and the continuum emission in the observed bands. In this case, an additional term of the variance is
\begin{equation}
\sigma^2_{S_L}|_{\rm stoc} = \frac{C^{uu}_\ell}{v_\ell C^{\delta \delta}_\ell}  \left [f(\nu_I) - f(\nu_V) \frac{g(\nu_I)}{g(\nu_V)} \right ]^2 (1 - r_M^2).
\end{equation}
If the Milky Way template traces the dust emission poorly $r_M \rightarrow 0$, this reverts to the full foreground contamination in Eq.\,\ref{eqn:totalerr}. Appendix\,\ref{app:lowrankgeo} describes the extension to many bands.

\section{Milky Way template cleaning applied to Planck x BOSS}
\label{sec:mwpxb}

The previous section argued that Milky Way templates can provide leverage to suppress foregrounds, especially in the case where the number of spectral channels is similar to the number of correlated continuum parameters. The analysis of Planck cross BOSS for \cii\ in \citep{2018MNRAS.478.1911P} is in this regime. This section considers the application of Milky Way templates in this case to boost signal to noise of the correlated line emission.

\subsection{Existing Milky Way templates}
\begin{figure} 
  \includegraphics[scale=0.45]{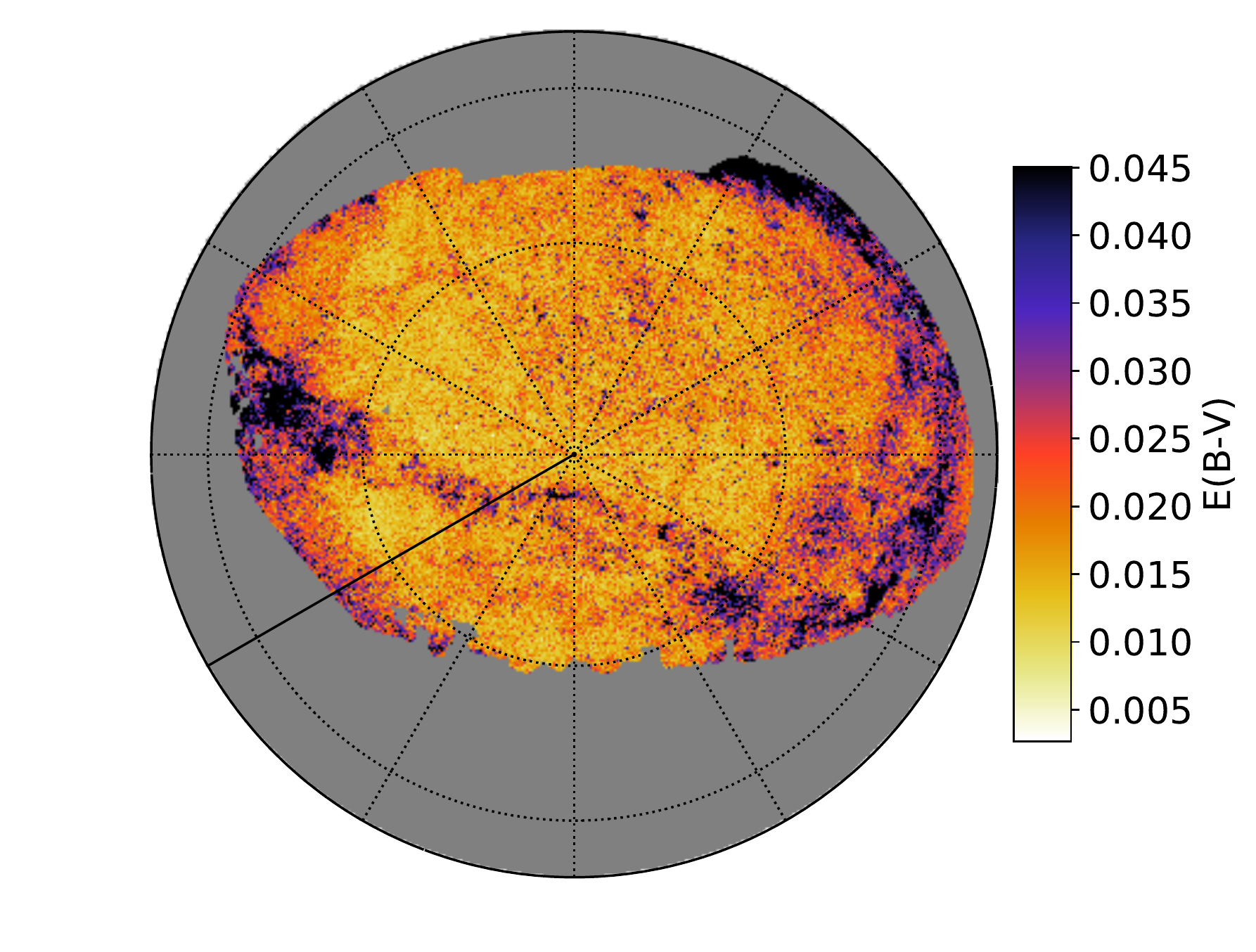}
	\caption{Reddening inferred by \citet{2015ApJ...810...25G} from stars in the Pan-STARRS survey in the BOSS-North quasar mask region.}
    \label{fig:greenmap}
\end{figure}

\begin{figure} 
  \includegraphics[scale=0.45]{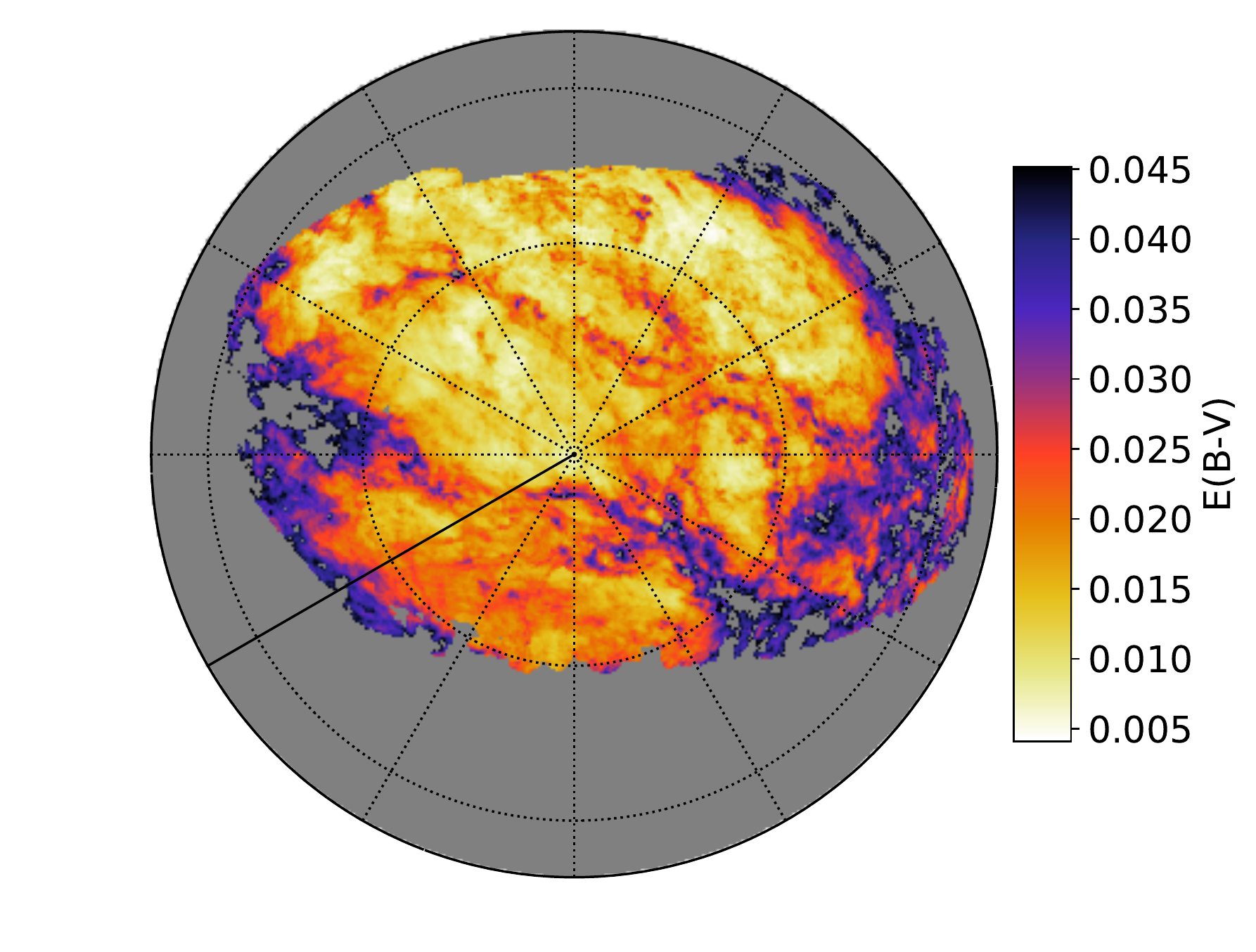}
	\caption{Reddening inferred by \citet{2017ApJ...846...38L} from HI tracers of dust in the BOSS-North quasar mask region.}
    \label{fig:lenzmap}
\end{figure}

\begin{figure} 
  \includegraphics[scale=0.55]{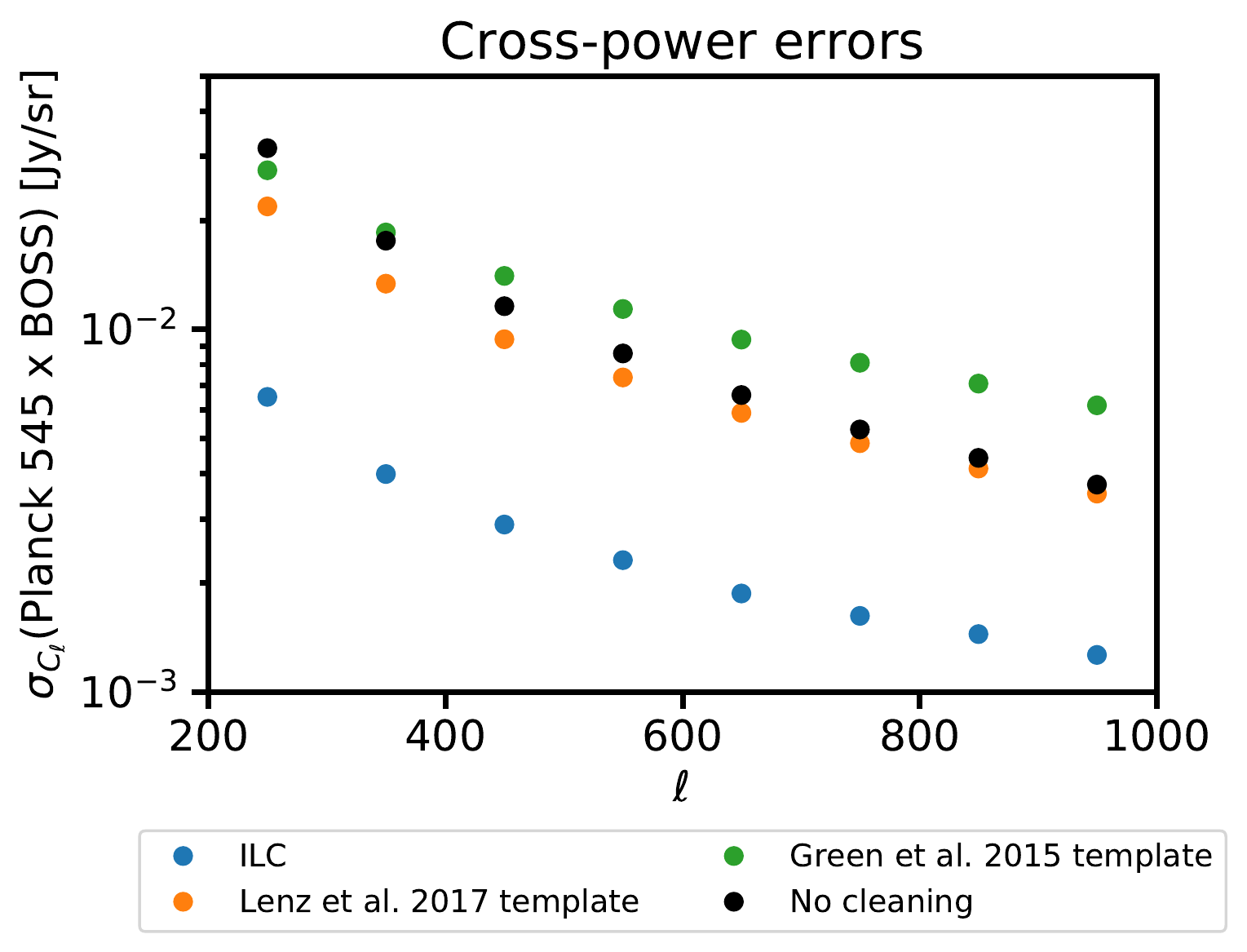}
  \caption{Bandpower errors for the Planck $545$\,GHz\,$\times$\,BOSS quasar overdensity with several linear-combination cleaning approaches. The ILC points show that a linear combination of Planck 353 and 857\,GHz with 545\,GHz removes much of the variance of the cross-power (but it is also modulates correlated continuum emission). The templates of the Milky Way from \citet{2015ApJ...810...25G} and \citet{2017ApJ...846...38L} modestly suppress errors on large scales, and \citet{2015ApJ...810...25G} increases errors $\ell > 300$ due to the noise in the determination of stellar reddening.}
  \label{fig:crosserrors}
\end{figure}

There are approaches to use broadband measurements to model thermal dust emission in the Milky Way \citep{2016A&A...596A.109P, 2014A&A...571A..11P, 2015ApJ...798...88M}. However, these galactic templates are prone to also contain extragalactic thermal dust emission, including correlated continuum emission. We instead seek models of thermal dust emission that do not rely on sensitivity to dust emission directly.

Thermal dust emission depends on both the column depth and the temperature. Existing inferences of temperature measure dust emission around the peak of the SED using broadband observations. These maps may, therefore, correlate with extragalactic radiation and cannot be used. Both reddening and HI abundance trace the dust column depth and can be inferred independently of extragalactic emission.

We use the direct inference of reddening to Pan-STARRS stars from \citet{2015ApJ...810...25G}, shown in Fig.\,\ref{fig:greenmap}. This map is noisy, capturing only the brightest features in the BOSS-North field. To reach higher sensitivity, we also consider the reddening map from \citet{2017ApJ...846...38L} inferred from HI4PI \citep{2016A&A...594A.116H} measurements of the HI column through $21$\,cm radiation (Fig.\,\ref{fig:lenzmap}). Both \citet{2015ApJ...810...25G} and \citet{2017ApJ...846...38L} are imperfect tracers of Milky Way dust emission. HI tracers for dust density become biased in high-density regions where ${\rm H}_2$ has formed \citep{2017ApJ...846...38L}, so the HI tracer is best in areas out of the galactic plane considered here. Reddening and HI are imperfect tracers of the dust column depth \citep{1989ApJ...345..245C, 2007ApJ...663..320F, 2016ApJ...821...78S}, and dust column is an imperfect tracer of the thermal dust emission.

Fig.\,\ref{fig:crosserrors} shows the reduction in bandpower errors when the \citet{2015ApJ...810...25G} and \citet{2017ApJ...846...38L} maps are projected out of the Planck 545\,GHz channel. This linear combination in map-space is only for visual demonstration. \citet{2017ApJ...846...38L} decreases variance on the largest scales, and \citet{2015ApJ...810...25G} adds noise for $\ell > 300$.

\subsection{Planck and BOSS data}

The Planck 353, 545 and 857\,GHz maps have full-width at half-max (FWHM) of 4.41, 4.47, and 4.23 arcmin, respectively, and are binned in HEALPix \citep{2005ApJ...622..759G} pixelization with $N_{\rm side} = 2048$. We use a point source mask that is the union of published masks \citep{2016A&A...594A..26P} across the three frequencies and apodized across $0.5^\circ$, and a galactic emission mask with $2^\circ$ apodization. In total, this leaves $f_{\rm sky} = 0.332$ in the BOSS survey region. Throughout, we will work in MJy/sr units, and convert between measurements in $K_{\rm CMB}$ units to MJy/sr using the mean coefficients 287.45 (353 GHz), 58.04 (545 GHz), and 2.27 (857 GHz) $({\rm MJy/sr}) / K_{\rm CMB}$ \citep{2014A&A...571A..30P}.

As a tracer of the underlying overdensity, we use the CORE uniform sample \citep{2012ApJS..199....3R} of the Baryon Oscillation Spectroscopic Survey (BOSS) \citep{2013AJ....145...10D} spectroscopic quasar sample \citep{2017A&A...597A..79P} from Data Release 12 (DR12) \citep{2015ApJS..219...12A}. We develop masks and overdensity as described in \citet{2018MNRAS.478.1911P}, identical to approaches for quasar clustering \citep{2012MNRAS.424..933W, 2014A&A...563A..54P, 2015MNRAS.453.2779E}. For DR12, this procedure yields a catalog of 178,622 quasars. Restricting to mask weight greater than $90\%$ gives 82,522 quasars over $8294\,{\rm deg}^2$ and an overlap with the Planck map of $6483\,{\rm deg}^2$ with 75,244 quasars. 

It is also beneficial to have a galaxy redshift sample that has no cross-correlation with the \cii\ emission, but does trace the CIB and SZ clustered emission, as part of a parametric model for the components. Here we use the BOSS DR12 CMASS luminous red galaxy (LRG) sample \citep{2017MNRAS.470.2617A, 2016MNRAS.455.1553R, 2015ApJS..219...12A}, which consists of 862,735 galaxies over $9376\,{\rm deg}^2$ and mean $z=0.57$, and is designed to be stellar-mass limited at $z > 0.45$. Sample selection here requires spectroscopic sectors \citep{2011ApJS..193...29A} with completeness $>70\%$ and redshift completeness $>80\%$. Additionally requiring $0.43 < z < 0.7$ and pixel coverage $>90\%$ gives 777,202 galaxies over $10,229\,{\rm deg}^2$. 

In both the quasar and LRG sample, we form the overdensity ${\boldsymbol \delta} = ({\bf n} - \bar {\bf n}) / \bar {\bf n}$ using systematic weights \citep{2014MNRAS.441...24A}. The LRG and CMASS mask regions are additionally apodized by a $0.5^\circ$ FWHM kernel. 

\subsection{Measured cross-powers}

\begin{figure} 
  \includegraphics[scale=0.35]{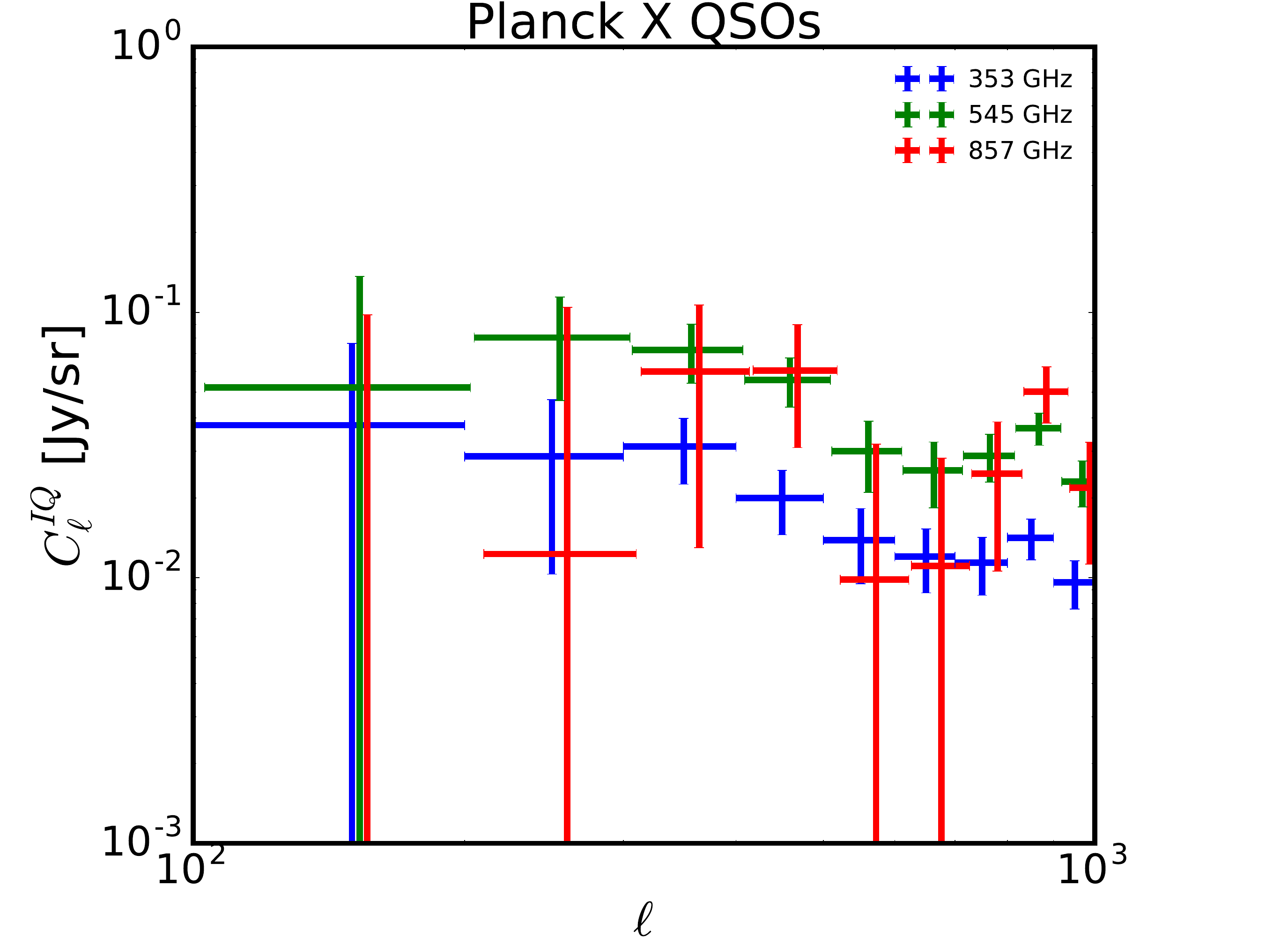}
  \includegraphics[scale=0.35]{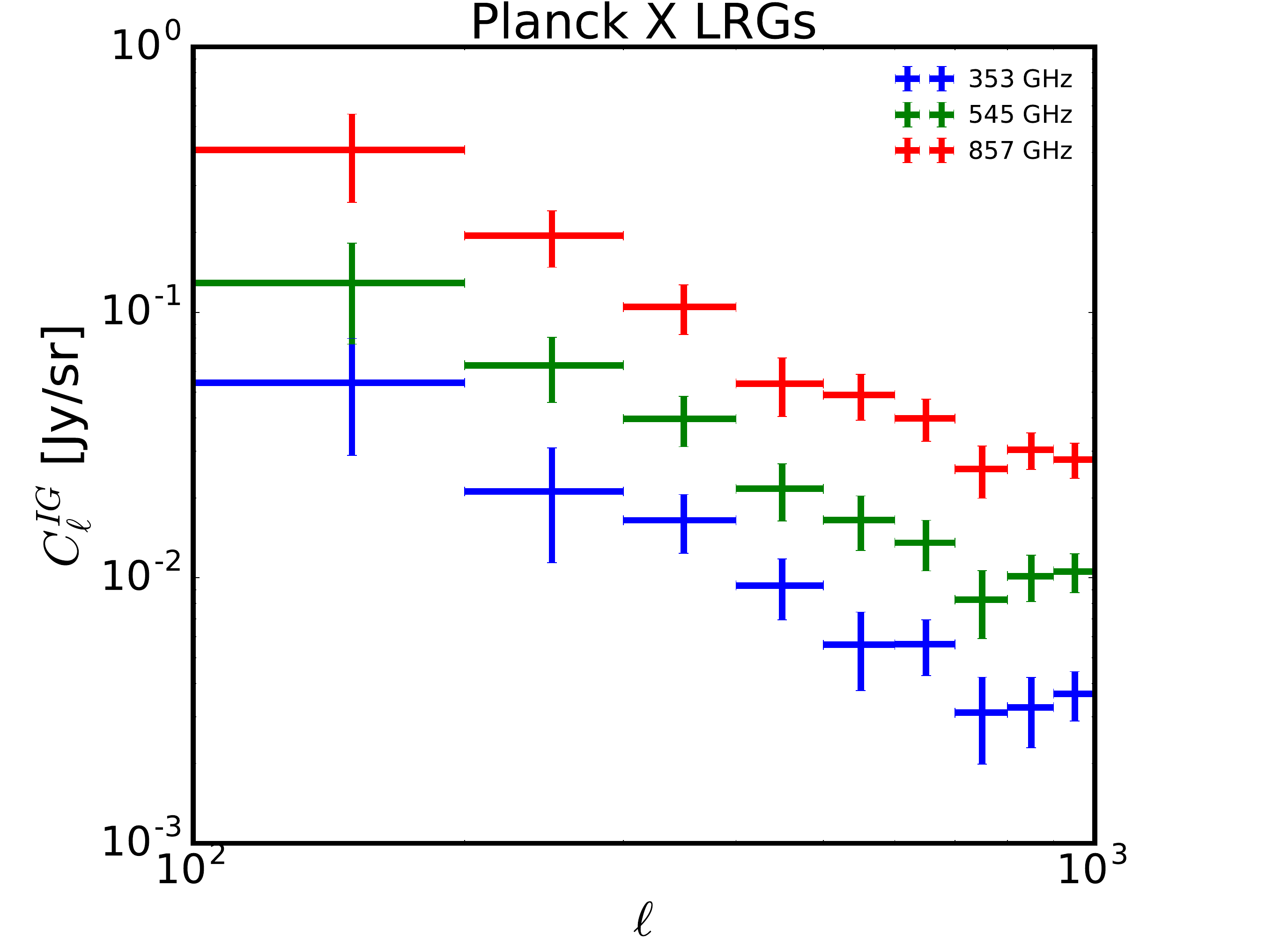}
  \caption{Measured angular cross-power spectra between Planck bands, quasars, and LRGs (1$\sigma$ errors indicated). 
  \label{fig:cl1}
}
\end{figure}

Fig.~\ref{fig:cl1} shows the six angular cross-powers used to estimate \cii\ emission and continuum nuisance parameters. These cross-spectra were performed over 3 Planck bands: 353, 545, and 857 GHz.  These maps were cross-correlated with overdensity maps of quasars and LRGs cataloged by the BOSS survey. We perform the cross-correlations over angular scales $\ell=100-1000$. These spectra are identical to those presented in \citet{2018MNRAS.478.1911P}. 

\begin{figure} 
  \includegraphics[scale=0.35]{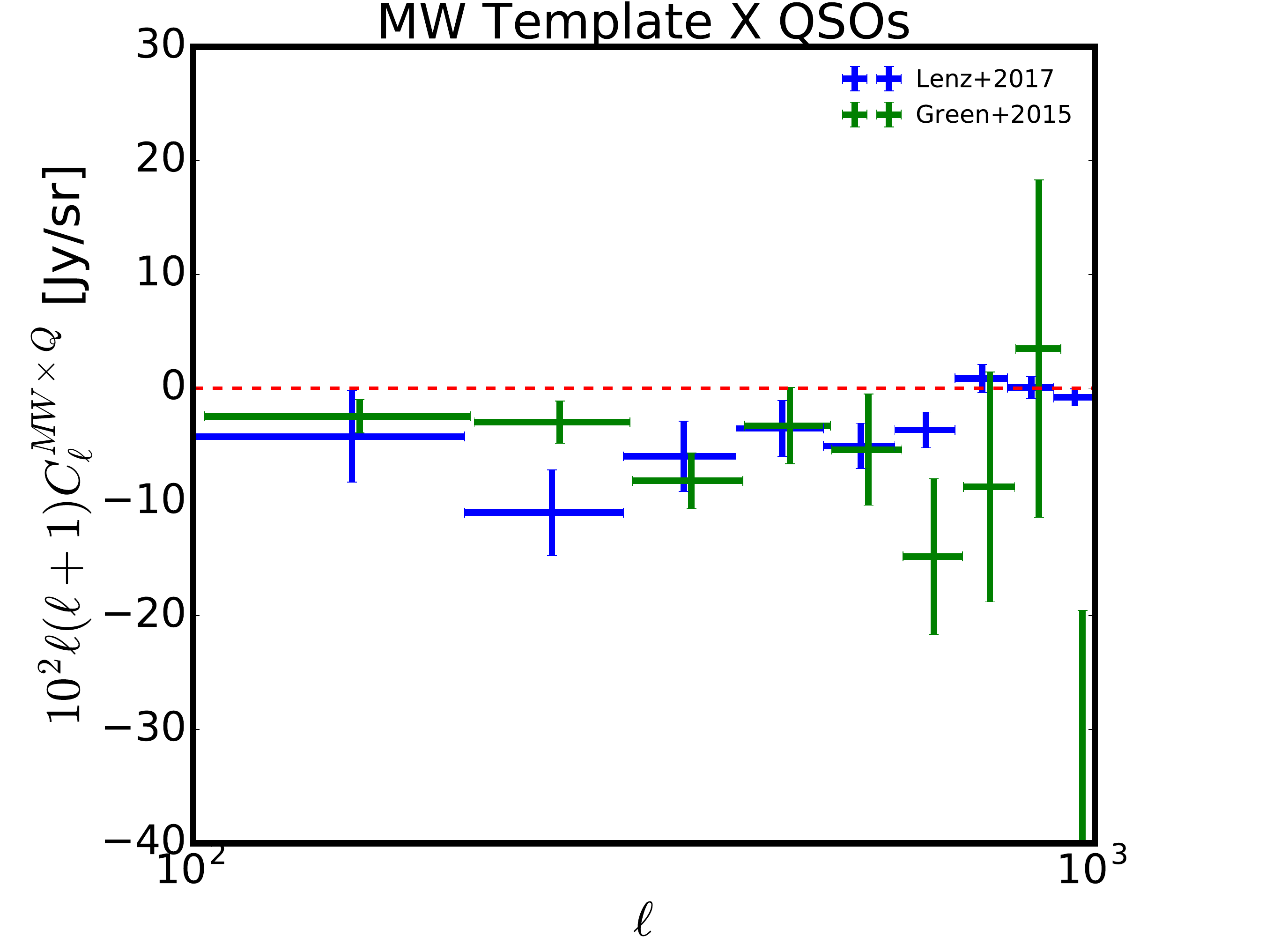}
  \includegraphics[scale=0.35]{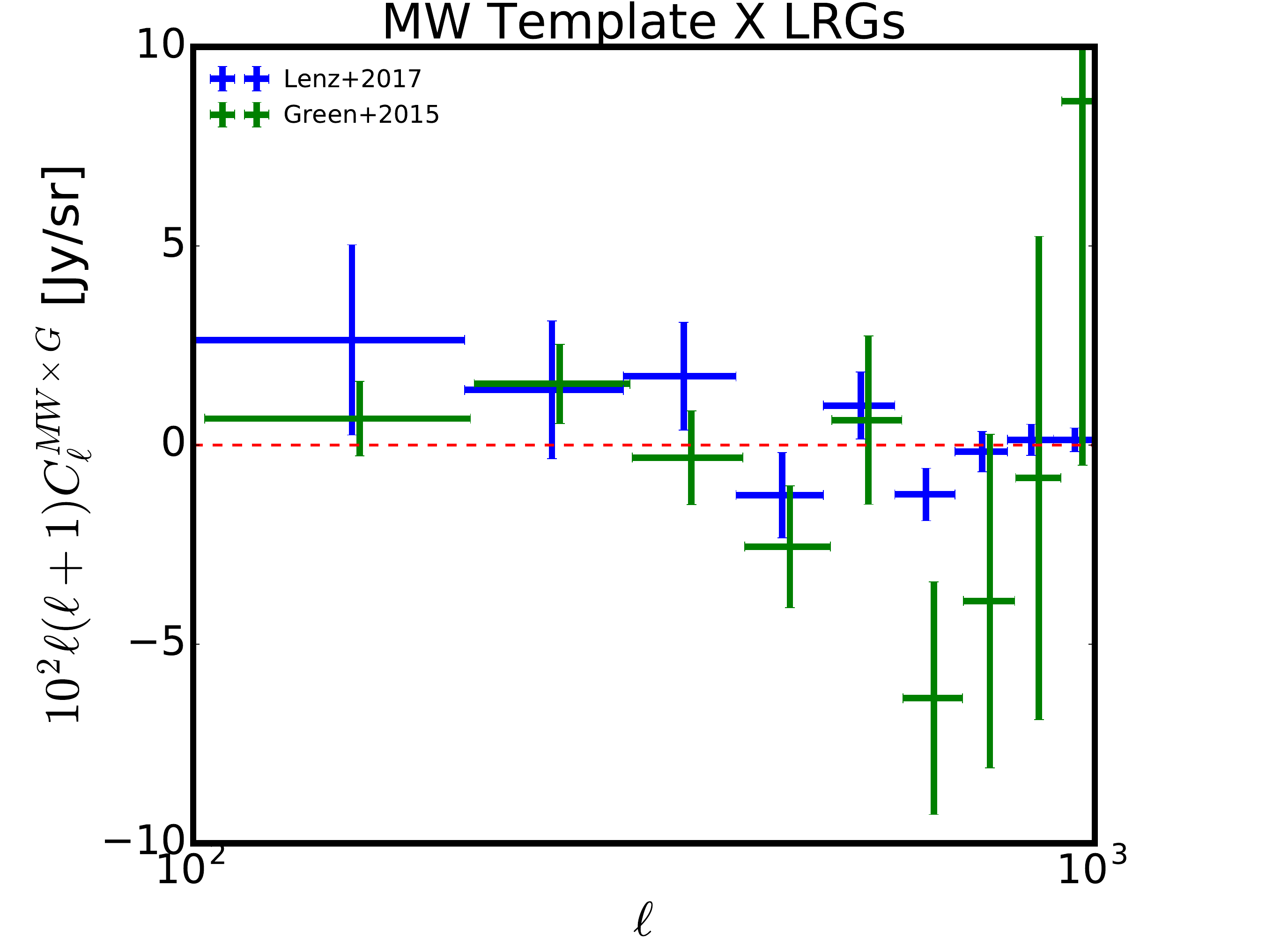}
  \caption{Angular cross-power spectra between the Milky Way templates from \citet{2015ApJ...810...25G} and \citet{2017ApJ...846...38L} for both the quasars and LRGs from the BOSS survey.  The \citet{2017ApJ...846...38L} map is a HI column density map scaled to the dust emission in the 545\,GHz Planck band.  The \citet{2015ApJ...810...25G} map is an extinction map multiplied by $10^6$. 
  \label{fig:dustcross}
}
\end{figure}

We also consider two Milky Way templates to marginalize over contamination in the Planck maps.  Specifically, we use the HI-derived dust map from \citet{2017ApJ...846...38L} and the reddening map from \citet{2015ApJ...810...25G}.  As observables, we cross-correlate these dust maps with both the quasar and LRG overdensity maps (Fig.~\ref{fig:dustcross}). Section\,\ref{ssec:cibciimodel} investigates evidence for correlation with large-scale structure.

\subsection{Models for CIB and CII signals}
\label{ssec:cibciimodel}

We briefly review the CIB and \cii\ emission models used in the \cii\ constraints from \citet{2018MNRAS.478.1911P}. We begin with the angular cross-power spectrum between CIB emission in a Planck map at frequency $\nu$ and a large-scale structure (LSS) tracer overdensity field, an expression given by
\begin{eqnarray} \label{E:cross}
C^{\rm LSS-CIB}_\ell &=& \int\frac{dz}{\chi(z)^2}\left(\frac{d\chi}{dz}\right)^{-1}b_{\rm LSS}b_{\rm CIB}(k,z)\nonumber\\
&&\times\frac{dN}{dz}(z)\frac{dS}{dz}(z,\nu)P_{\rm DM}\left(\frac{\ell}{\chi(z)},z\right)\, ,
\end{eqnarray}
where $\chi$ is the comoving distance, $dN/dz$ is the redshift distribution of the LSS tracer, $P_{\rm DM}(k,z)$ is the matter power spectrum computed using CAMB, and $b_{\rm LSS}$ and $b_{\rm CIB}(k,z)$ are the clustering biases for the LSS tracer and the CIB emitters, respectively.  For the LSS tracers we consider, we fix $b_{\rm LRG}=2.1$ and we allow $b_{\rm QSO}$ to float in the range 3.2--3.8.

The redshift source distribution of CIB sources can be written as
\begin{eqnarray}
\frac{dS_{\nu}}{dz} = \frac{c}{H(z)(1+z)}\int\,dL\frac{dn}{dL}(M,z)\frac{L_{\nu(1+z)}}{4\pi}\, ,
\end{eqnarray}
where $dn/dL$ is the infrared galaxy luminosity function and $\mathrm{L_{\nu(1+z)}}$ is a model for the CIB luminosity emitted at rest frame frequency $\nu(1+z)$.  $b_{\rm CIB}(k,z)$ and $dn/dL$ are predicted using the same halo model as in \citet{2012MNRAS.421.2832S}.  The model for the luminosity depends on several parameters; however we only allow 3 parameters in our fit to vary: a luminosity amplitude $L_0$, a redshift evolution parameter $\delta$, and the dust temperature $T_d$.  

Following \citet{2018MNRAS.478.1911P} we extend the CIB spectrum model in \citet{2012MNRAS.421.2832S} to include \cii\ emission in the host galaxies, with amplitude $A_{\rm CII}$. We also add correlated thermal SZ emission, which we model with an amplitude $A_{tSZ}$ multiplying a spectral template.  See \citet{2018MNRAS.478.1911P} for more details.  

The bandpower covariance that appears in the likelihood is described in \citet{2018MNRAS.478.1911P} and includes $\ell$-$\ell'$ correlations from masked sky regions. 

\subsection{CIB and CII parameter estimates}
\label{ssec:mcmcparam}

\begin{figure*} 
  \includegraphics[scale=0.55]{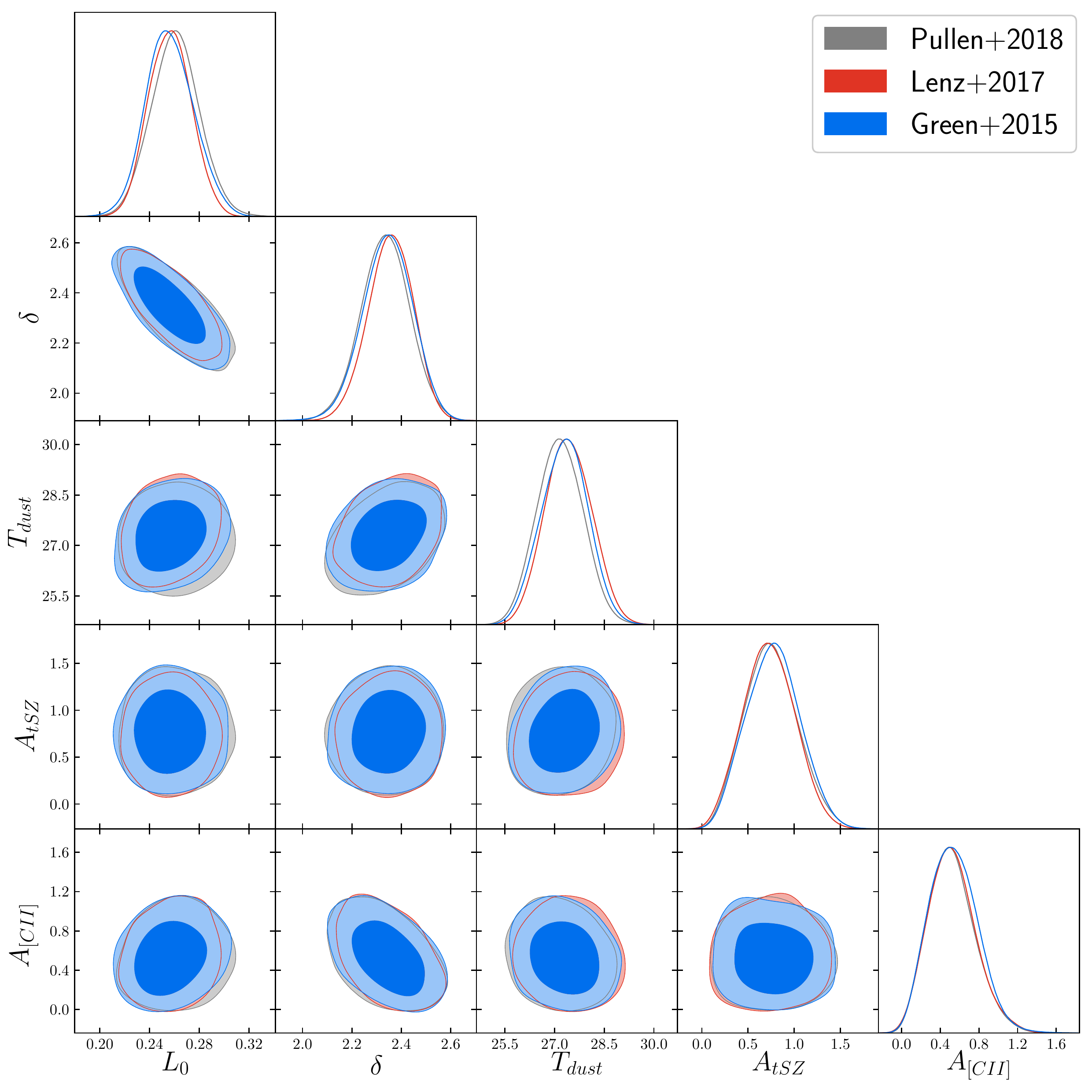}
  \caption{68\% and 95\% posteriors for the CIB, $A_{tSZ}$ and $A_{CII}$ parameters from the MCMC. The inclusion of Milky Way dust emission templates from \citet{2015ApJ...810...25G} and \citet{2017ApJ...846...38L} does not aid detection of \cii\ in Planck $\times$ BOSS quasars.
  \label{fig:mcmc}
}
\end{figure*}

We perform our MCMC analysis using CosmoMC \citep{2002PhRvD..66j3511L} for the 6 parameters
\begin{eqnarray}
\Xi \equiv \{T_d, \delta, L_{\mathrm{0}}, A_{\rm CII}, A_{\rm tSZ}, b_{\rm QSO} \}.
\end{eqnarray}
As in \citet{2018MNRAS.478.1911P}, we fit the parameters using the 6 cross-power spectra between the 3 high-frequency Planck bands and the quasars and LRGs presented in Fig.~\ref{fig:cl1}; these are measured in 9 bins over $100<\ell<1000$.  We label this fit the ``Pullen+2018'' fit.  We include the full covariance matrix between all the cross-power spectra. Also as before, we fit the mean levels of the CIB emission in the 3 Planck bands \citep{2012A&A...542A..58B} as well as 10 star formation rate density measurements \citep{2014ARA&A..52..415M}. 

The Milky Way templates show some evidence of correlation with the quasar overdensity. This could bias the determination of the correlated \cii\ amplitude, so we additionally marginalize over a nuisance clustering anisotropy based on the CIB$\times$quasar clustering model with amplitude $\alpha$.

Fig.~\ref{fig:mcmc} shows the parameter fits before and after including Milky Way emission templates.  The posteriors for CIB parameters and thermal SZ amplitude do not change significantly from \citet{2018MNRAS.478.1911P}.  The posterior distribution of $A_{\rm CII}$ is only modestly impacted by marginalizing over the \citet{2017ApJ...846...38L} template.  We also find that the $A_{CII}$ measurement is not degenerate with $\alpha$ for either Milky Way template.  The MCMC constrains $-0.092<\alpha_{\rm Lenz}<-0.025$ and $-0.121<\alpha_{\rm Green}<-0.042$ at 95\% confidence. This indication of correlation between the Milky Way-only tracers and the quasar overdensity warrants future investigation. Synchrotron emission from the quasars could correlate with the anisotropy of the synchrotron background in the $21$\,cm maps. 

\subsection{Interpreting the MCMC results}
\label{ssec:interpmcmc}

\begin{figure} 
  \includegraphics[scale=0.55]{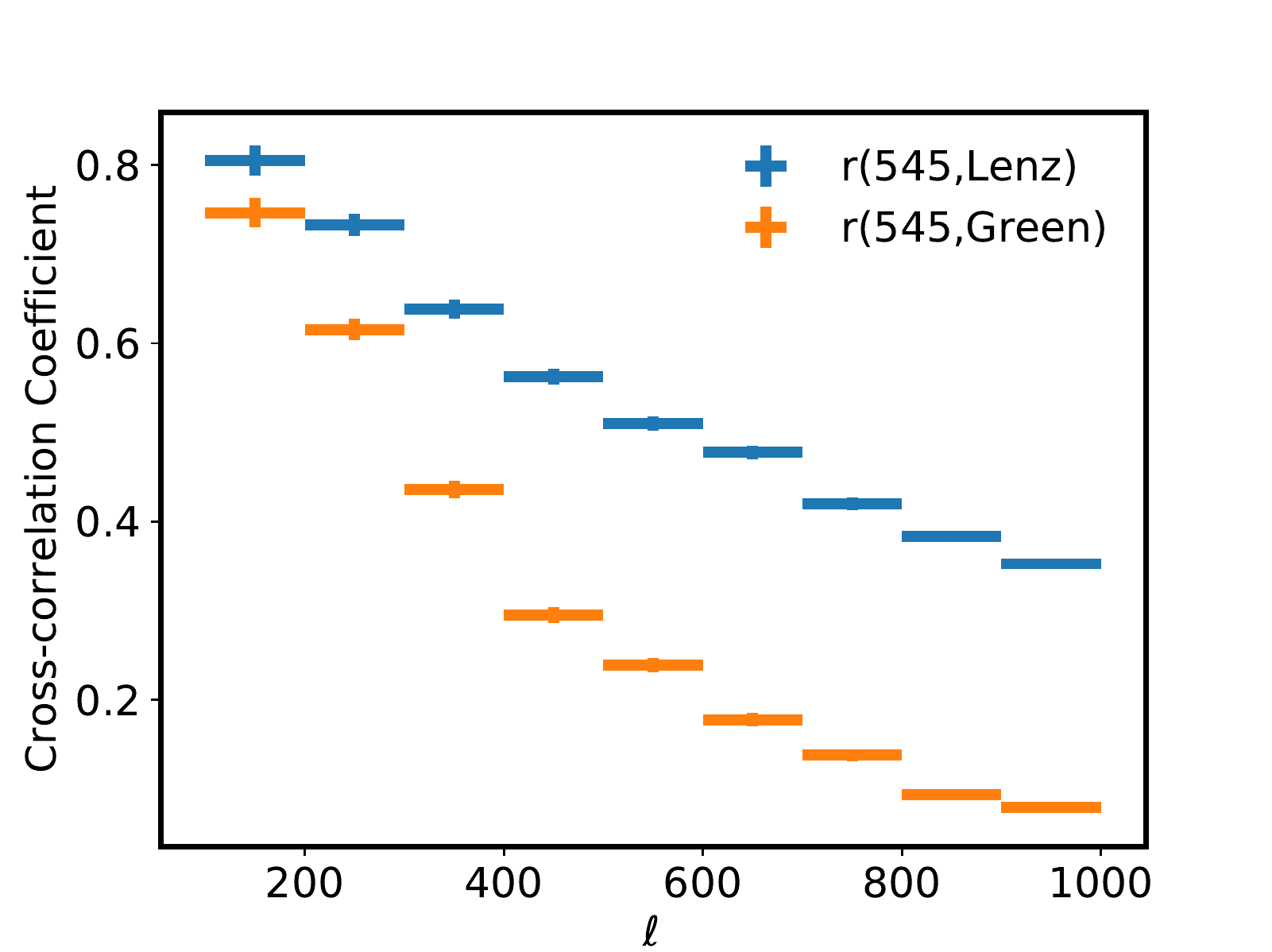}
  \caption{Stochasticity between the \citet{2017ApJ...846...38L} HI map and Planck 545\,GHz defined as $r_{M, {\rm Lenz}} = C_\ell({\rm Lenz} \times 545) / \sqrt{C_\ell({\rm Lenz} \times {\rm Lenz}) C_\ell(545 \times 545)}$.  We also plot the same for the \citet{2015ApJ...810...25G} reddening map. Large-scale features in the dust emission are traced well by HI and reddening, but smaller scales become compromised by dust temperature variations, molecular gas, and noise.}
  \label{fig:cross_tlg}
\end{figure}

\begin{figure} 
  \includegraphics[scale=0.55]{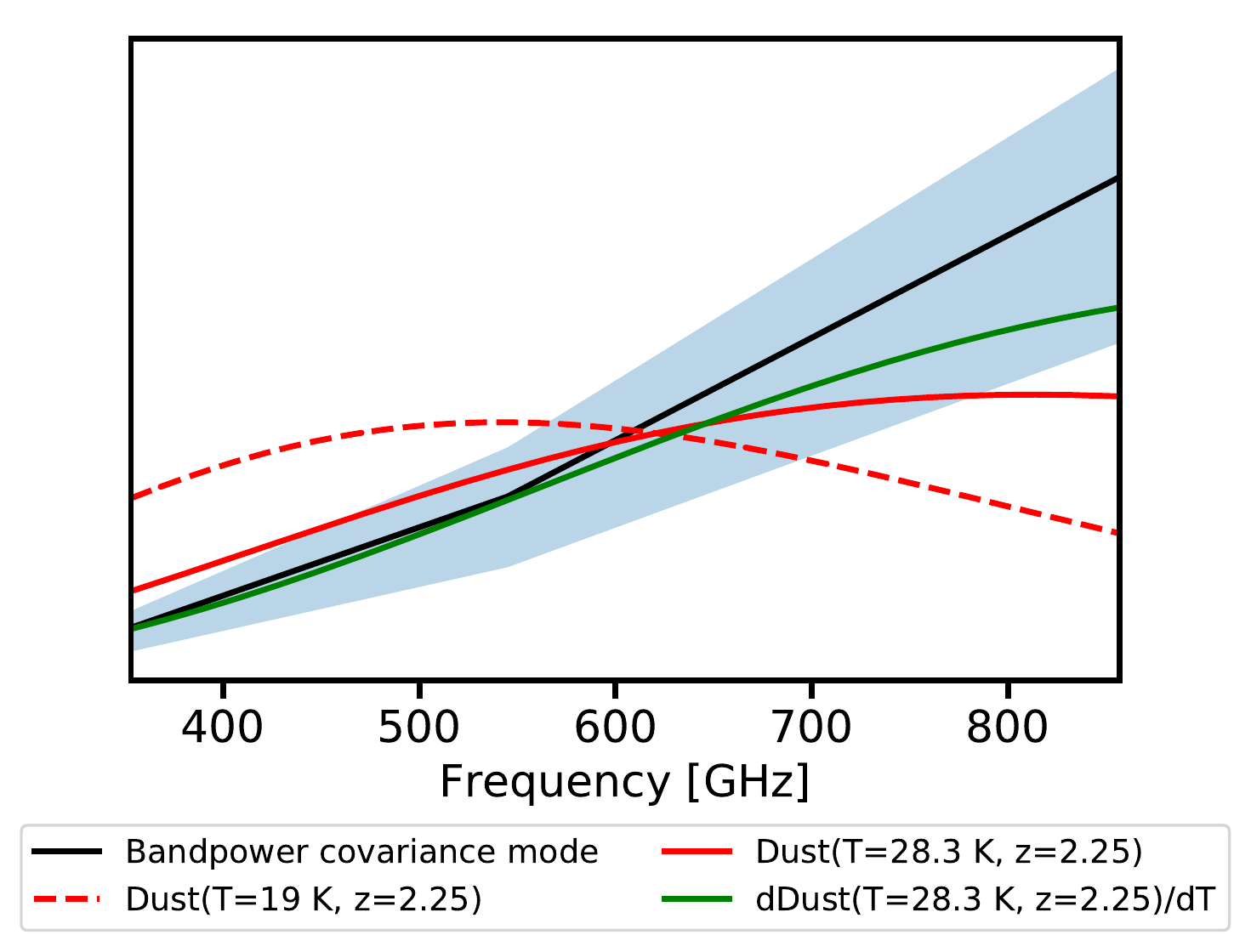}
	\caption{Principal mode of the Planck $\{353, 545, 857\}\,{\rm GHz} \times$\,BOSS-quasar bandpower covariance in frequency space. The blue range shows variation of this mode across $200 < \ell < 1000$. The principal mode is consistent with thermal dust emission at $19$\,K with $\beta=1.5$ at $z=0$, from the Milky Way. The colored curves show linear modes of a correlated continuum model which approximates the behavior of the nonlinear parameter estimate in Section\,\ref{ssec:mcmcparam}. The correlated continuum SED is marginalized over two modes shown here at $27.2$\,K, $\beta=1.5$ and $z=2.25$, where the derivative mode provides a degree of freedom to constrain the temperature. Marginalizing over the correlated continuum suppresses the principal variance of the Milky Way contamination, due to the coincidental similarity in SED. Extragalactic emissions from $z=2.25$ and $19$\,K is also shown as an example of a poorer SED match.}
    \label{fig:covariance_mode}
\end{figure}

We can interpret the lack of impact of Milky Way templates in Planck$\times$BOSS-quasar measurements using the multi-band formalism in Sec.\,\ref{sec:manybands}. To do this, recast relevant parts of the full nonlinear model as linear modes that fit the correlated continuum emission as in Eq.\,\ref{eqn:multierr}. Let the data vector be the cross-powers of the Planck 353, 545 and 857\,GHz maps and BOSS quasars. Take a simple model for the SED of the correlated continuum with $I^{\rm dust}_\nu = \nu^\beta B[T_{\rm dust}, (1+z) \nu]$ where $B(T_{\rm dust}, \nu)$ is the Planck law, and we fix $\beta=1.5$. To build a linear model, let ${\bf M}$ have a mode which is the best-fit dust with $T_{\rm dust} = 27.2$\,K, evaluated at $z=2.25$ (near the peak of the quasar number density). To model changes in temperature, also add a mode $d I^{\rm dust}_\nu / d T(\nu)$.

For the covariance in Eq.\,\ref{eqn:multierr}, use the bandpower covariance of the Planck data. Fig.\,\ref{fig:covariance_mode} shows this principal mode of the bandpower covariance, which is consistent with emission from the Milky Way ($\beta=1.5$, $T_{\rm dust} = 19$\,K from $z=0$). The principal mode is spectrally similar to emission with $T_{\rm dust} = 27.2$\,K from $z=2.25$ in the correlated continuum model. The similarity in spectral indices means that the operation that marginalizes over correlated continuum emission is also very effective at suppressing the Milky Way emission. In the setting of Eq.\,\ref{eqn:totalwtemplate}, the correlated and uncorrelated continuum terms have similar SED so $f(\nu) \approx g(\nu)$ and the second term of the variance is highly suppressed. The value of a Milky Way template map is diminished in this case.

In the linear model, the \citet{2017ApJ...846...38L} template increases the signal-to-noise in the line amplitude estimate by only $8\%$. Adding a derivative mode $d I^{\rm dust}_\nu / d T$ (as a linearized temperature parameter) reduces the improvement of the \citet{2017ApJ...846...38L} template to $3\%$. Most of this improvement comes from $\ell < 500$, where the Milky Way template is best correlated to Planck 545\,GHz (Fig.\,\ref{fig:cross_tlg}). In contrast, if the extragalactic dust temperature were the same as the Milky Way's $\approx 19$\,K (but redshifted to $z=2.25$), the spectral match between the correlated SED modes and the galactic dust would not be nearly as good. For a $19$\,K emitter at $z=2.25$, the dust spectrum crests in the $545$\,GHz band, while at $z=0$, the dust emission in $857$\,GHz is brighter. In this scenario, the \citet{2017ApJ...846...38L} template would lead to a $27\%$ improvement in the line constraint. Hence, the Planck $\times$ BOSS quasar measurements benefit from a lucky coincidence between the galactic and redshifted extragalactic emission SED. This similarity allows the marginalization over the extragalactic SED to also effectively deweight the Milky Way emission.

\section{A true tomographic survey}
\label{sec:tomographic}

Section\,\ref{sec:manybands} argued that an intensity survey needs at least an many frequencies as degrees of freedom in the correlated continuum mode, and argued that suppression of continuum contamination is equivalent to throwing out $k_\parallel$ (line of sight wavevector) information. This section describes how correlated continuum emission appears in the full 3D power spectrum $P_\times (k_\perp, k_\parallel)$ of tomographic surveys with numerous spectral bins. 

Here, we will mock up a future tomographic survey to show the correlated continuum and line emission in the full 3D power spectrum $P_\times (k_\perp, k_\parallel)$ for wavenumbers perpendicular $k_\perp$ and parallel $k_\parallel$ to the line of sight. Correlated continuum emission enters at low $k_\parallel$. Cutting $k_\parallel < k_\parallel^{\rm cut}$ recovers the line correlation independently of the continuum but results in some loss of sensitivity at low $|k|$. 

\subsection{Impact of correlated continuum emission in the 3D cross-power}

Before doing a numerical simulation, we can get intuition for the impact of the SED on the intensity mapping cross-power by working in the Hubble approximation. Here the SED is a direct convolution of the overdensity field in the line of sight direction, which becomes a multiplication in $k_\parallel$ space.

The surface brightness at frequency $\nu$ is the integral of comoving specific emission intensity $j(\nu,z)$ across all redshifts in a tomographic survey slice
\begin{equation}
I_\nu = \int \frac{dz}{1+z} \frac{d \chi}{dz} j(\nu, z),
\label{eqn:sbintegral}
\end{equation}
where $\chi(z)$ is the comoving distance. For a thin tomographic redshift slice $\delta z$, the brightness of the line emission will not depend on $\delta z$. In contrast, the correlated component of the continuum emission will be the integral in Eq.\,\ref{eqn:sbintegral} over the redshift slice, and so it will depend linearly on the redshift width $\delta z$. 

Let the continuum SED at frequency $\nu$ from galaxies at redshift $z$ be $\Theta[\nu(1+z)]$, and take a greybody dust \citep{2002PhR...369..111B}. Evaluate the SED on the comoving grid points and find its discrete Fourier transform $\tilde \Theta(k_\parallel)$, normalized such that $\tilde \Theta(k_\parallel=0) = 1$. If the overdensity in 3D k-space is ${\boldsymbol \delta}_k$, then the convolution with the full line plus continuum SED is
\begin{equation}
{\boldsymbol \delta}_k^{IM} = S_L \left [ 1 +\Delta zf_{\rm cont} \tilde \Theta(k_\parallel) \right ] {\boldsymbol \delta}_k,
\end{equation}
where $\Delta z$ is the total redshift range, or $N_{\rm chan} \delta z$ for $N_{\rm chan}$ channels, and $f_{\rm cont}$ is the fraction of continuum to line intensity from galaxies per unit redshift. Relating this to the Planck-BOSS cross correlation model in Section\,\ref{ssec:cibciimodel}, $f_{\rm cont} = \frac{2.5}{A_{\rm CII}}$. Here $S_L$ is the line brightness times cosmological linear bias $I_{\rm CII} b_{\rm CII}$, and we assume that the dust emission shares the same bias for simplicity.

The cross-power ${\boldsymbol \delta}_k^{IM} {\boldsymbol \delta}_k^{g*}$ binned onto $k_\parallel$ and $k_\perp$ is then
\begin{equation}
P_\times (k_\perp, k_\parallel) = S_L b_g \left [ 1 + \Delta zf_{\rm cont} \tilde \Theta(k_\parallel) \right ] P_{\delta \delta}(k_\perp, k_\parallel),
\label{eqn:cross_power_cont}
\end{equation}
where $P_{\delta \delta}(k_\perp, k_\parallel)$ is the power spectrum of the underlying dark matter field ${\boldsymbol \delta}$. This is just the expected intensity mapping cross-power $P_\times (k_\perp, k_\parallel) = S_L b_g P_{\delta \delta}(k_\perp, k_\parallel)$ plus a term that multiplies the $k_\parallel$ direction. In the case where galaxies have a constant SED over this spectral range, $\Theta(k_\parallel)$ is a $\delta$-function, and the continuum only appears at $k_\parallel = 0$. In practice, continuum emission will contaminate higher $k_\parallel$ because the SED has additional spectral structure, the Hubble approximation will break down, and there may be mixing between $k_\parallel$ modes because of the survey geometry.

\begin{figure}
  \includegraphics[scale=0.5]{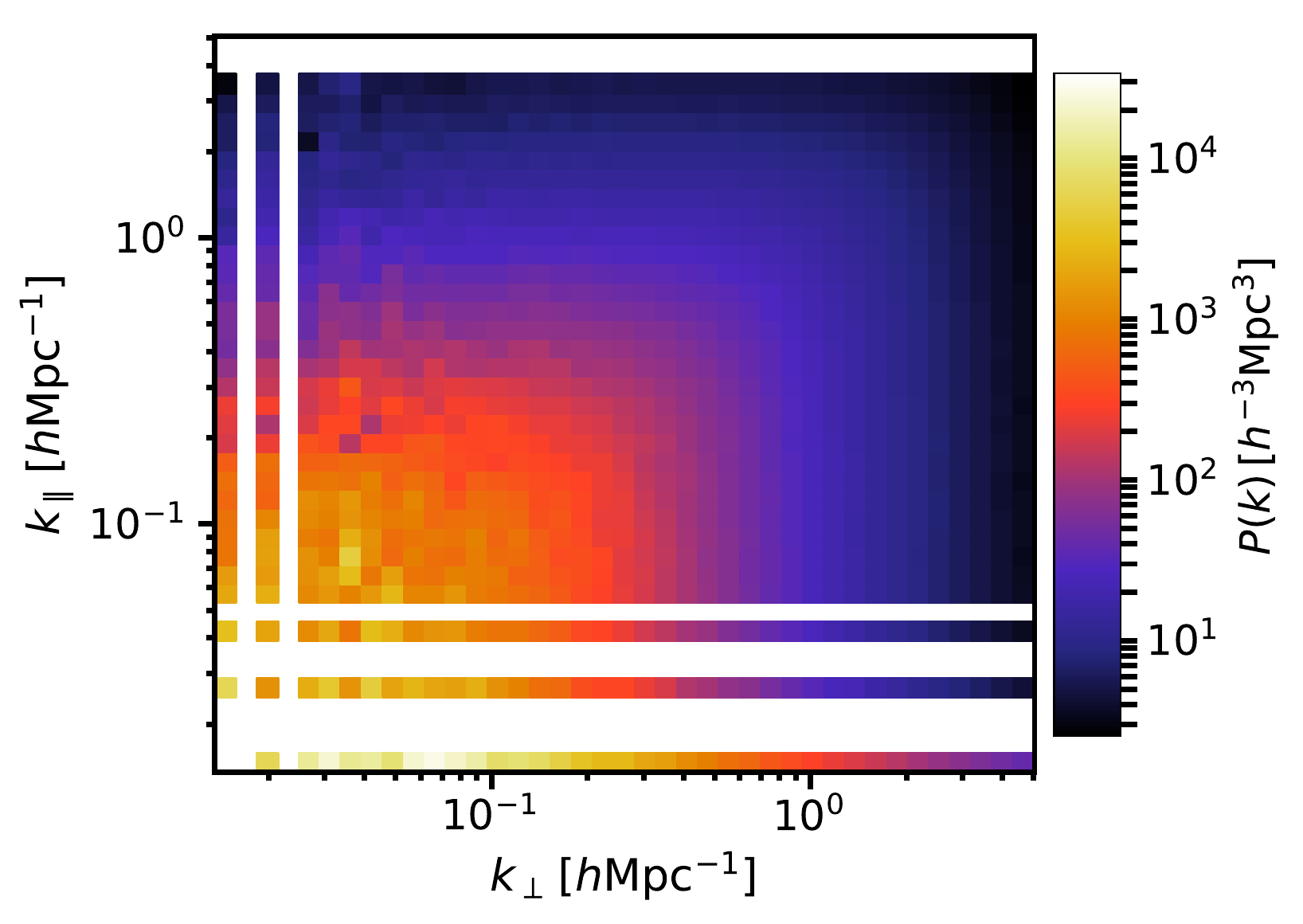}
  \caption{Simulated 2-dimensional cross-power between a galaxy redshift survey and a tomographic intensity survey with both correlated line and continuum emission. The contamination from the continuum is visible in the high amplitude of the lowest two $k_{\parallel}$ modes. This calculation uses a Blackman window to apodize in the frequency direction. At higher $k_{\parallel}$, one can see from the symmetry of the power spectrum (we do not simulate redshift space distortions) that continuum contamination is negligible. Note that the $k_{\parallel}=0$ mode is not shown on this log plot. \label{fig:2d_pk_sim}}
\end{figure}

\begin{figure}
  \includegraphics[scale=0.5]{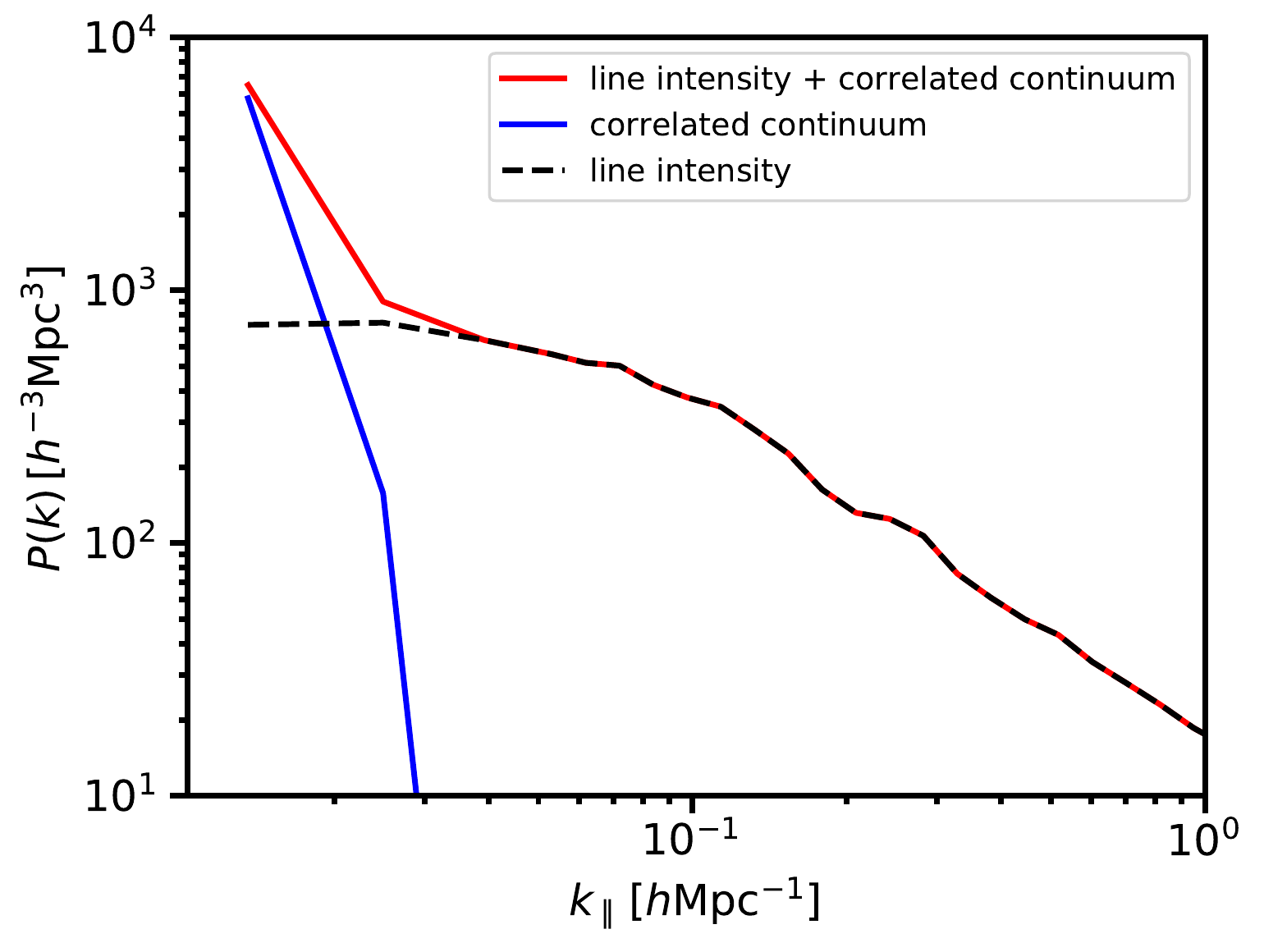}
  \caption{The cross-power of correlated line, continuum and line+continuum emission with cosmological overdensity as a function of $k_{\parallel}$ after averaging each 2-dimensional power spectrum over $k_{\perp}$. The continuum emission only contaminates the two lowest $k_{\parallel}$ modes, even when the assumed \cii\ amplitude is 20 times lower than \citet{2018MNRAS.478.1911P}. These calculations use a Blackman window to apodize in the frequency direction.
  \label{fig:pk_avg_sim}}
\end{figure}

We simulate the effect of correlated continuum emission on the cross-power spectrum by drawing Gaussian sample maps from a $z=2.25$ power spectrum \citep{2013JCAP...11..044D}. The maps are 500$\,{\rm Mpc}/h$ cubes with a pixel size of 1$\,{\rm Mpc}/h$, which corresponds to $2<z<2.5$ in the line-of-sight direction. The continuum SED is then simulated by adding a greybody dust SED whose magnitude is proportional to the overdensity at each voxel, and redshifted according to that voxel's redshift. To illustrate the sensitivity of a future survey, we take $A_{\rm CII}$ a factor of $20$ smaller than the posterior in Section\,\ref{ssec:mcmcparam}. Since our focus is only on the ratio of correlated continuum to line emission, we do not apply the multiplicative prefactor $S_L b_g$ (Eq.\,\ref{eqn:cross_power_cont}) to our simulated intensity maps. Therefore, the units of the power spectra in Figs.\,\ref{fig:2d_pk_sim} and \ref{fig:pk_avg_sim} are ${\rm Mpc}^3h^{-3}$.  Fig.\,\ref{fig:2d_pk_sim} shows the 2-dimensional power spectrum of one realization of this simulated model. (The impact of the continuum convolution is partly stochastic, so averaging over many simulations also averages down part of the correlated continuum.)

\begin{figure}
  \includegraphics[scale=0.5]{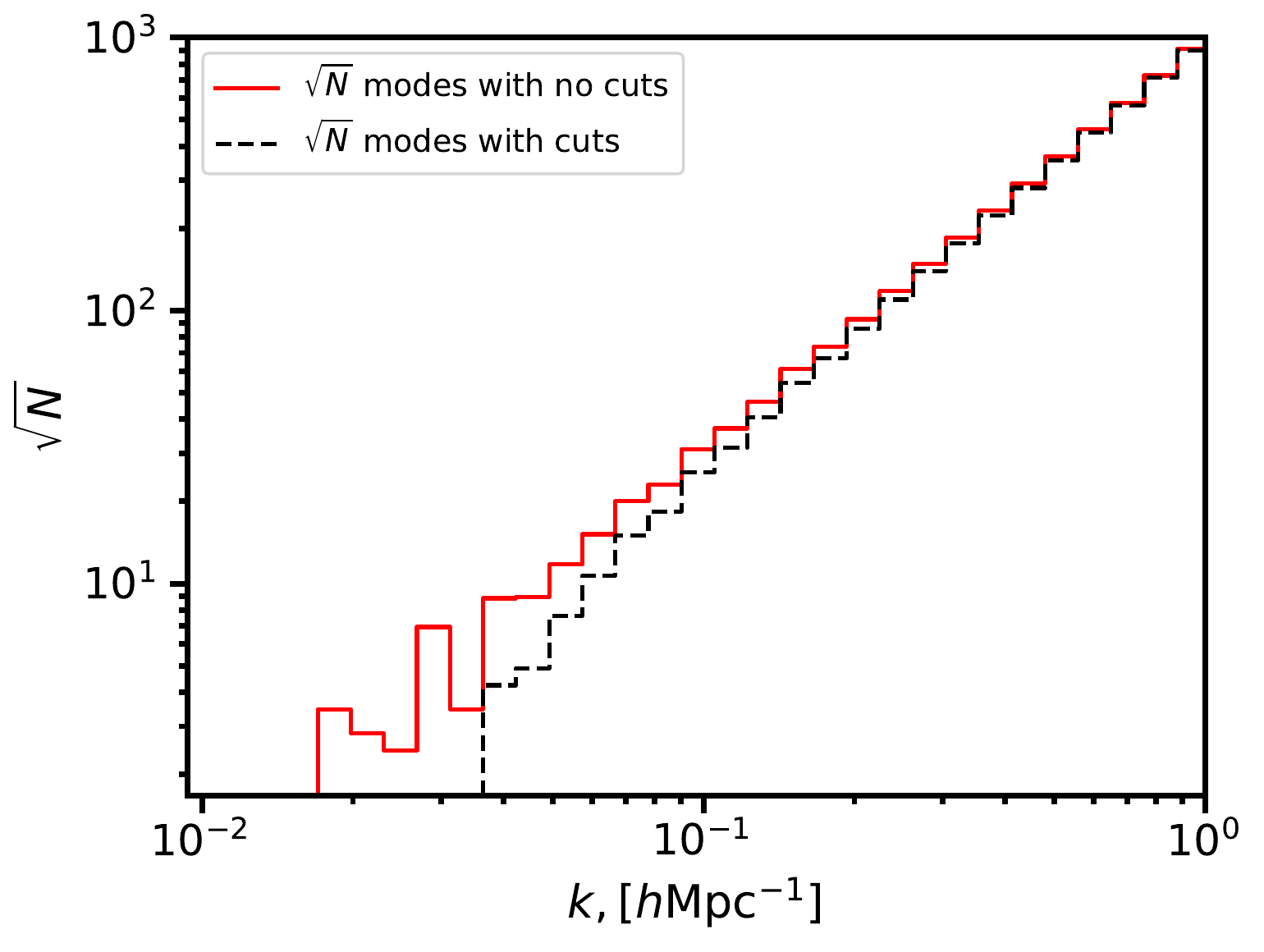}
  \caption{Decrease in number of modes available to estimate the $|k|$ bandpower due to cutting the two lowest $k_{\parallel}$ bins in Fig.\,\ref{fig:2d_pk_sim}. The ratio of the two curves is the fractional error increase.
  \label{fig:sim_modes_kept}}
\end{figure}

Only the two lowest non-zero $k_{\parallel}$ modes are noticeably biased by the continuum SED. This is shown more clearly in figure \ref{fig:pk_avg_sim}, in which we plot the power spectrum, averaged over $k_{\perp}$, as a function of $k_{\parallel}$. Figure \ref{fig:sim_modes_kept} shows the decrease in the number of modes for $|k|$ by cutting the two lowest non-zero $k_{\parallel}$ modes from the binned one-dimensional power spectrum.

We can also revisit the question from Sec.\,\ref{ssec:howmany} of the number of bands that a survey needs to treat correlated continuum emission without significant signal loss. In the tomographic case, the number of $k_\parallel$ modes corresponds roughly to the number of frequency channels in the spectrometer. Removing the 5 lowest $k_\parallel$ bins rejects the majority of correlated continuum. Schematically, if there are a total of 500 channels, this results in a $5/500 = 1\%$ loss in signal. Conversely, if there are 10 channels and 5 $k_\parallel$ modes must be removed, $50\%$ of the signal is lost. Calculations here have made simple assumptions for the number of degrees of freedom in the correlated continuum SED. In practice, both the correlated and uncorrelated continuum may be more complex through their astrophysics, or through interaction with the instrument.

\section{Discussion}
\label{sec:discussion}

Cross-correlation between intensity mapping datasets and galaxy redshift surveys has become a gold-standard for robustness against foreground contamination. This cross-correlation is not only sensitive to the line emission in target galaxies, but also the complete, correlated SED. The inference of line brightness needs to marginalize over non-line contributions in the SED. The influence of correlated continuum emission has only recently been appreciated \citep{2014A&A...570A..98S, 2017ApJ...838...82S, 2018MNRAS.478.1911P, 2018arXiv181000885C}. Here we have expanded the consideration to a more general survey to understand circumstances when correlated continuum emission can be a significant pitfall for intensity mapping. 

Correlated continuum emission should be estimated jointly (self-consistently) with bright Milky Way foreground cleaning. For most intensity mapping surveys, a synchrotron or dust continuum dominate the correlated SED. This SED is similar, but not identical to the bright continuum emission from the Milky Way. We show that a properly-formed likelihood will deweight foregrounds in $C_\ell$-space analogously to map-space cleaning through a linear combination. Doing the deweighting as part of the parameter estimation from the power spectra self-consistently treats the uncorrelated and correlated continuum emission. In contrast, an analysis that cleans in map-space before the likelihood analysis requires simulations to quantify residual correlated continuum emission and signal loss. 

Intensity mapping observations with a limited number of bands have unique challenges that we consider in the case of Planck cross BOSS \citet{2018MNRAS.478.1911P}. When the number of channels is approximately the number of correlated continuum parameters, there is limited latitude to 1) deweight the bright uncorrelated foregrounds while also marginalizing over the correlated continuum, 2) verify that the line amplitude estimate is not biased because of an incomplete continuum model.

We reconsider the analysis of Planck cross BOSS with the addition of Milky Way-only templates. These templates have the potential to provide leverage to separate and clean Milky Way emission independently of the correlated dust continuum. However, in practice we find that existing templates \citep{2015ApJ...810...25G, 2017ApJ...846...38L}, do not yet provide significant benefit, due both to the stochasticity with which they trace the Milky Way and to a similarity in the correlated and Milky Way dust continuum spectra.

Our analysis of Planck cross BOSS will have shortcomings for future intensity mapping data, which move from simple detection to astrophysical characterization. Rather than taking a parametric model for the clustered emission, future work could directly solve for the $SED(z_i, \nu_j)$ and anisotropy structure $C_\ell$ of the correlation, similarly to a ``CMB-only'' approach \citep{2013JCAP...07..025D}. Direct inference of the clustering anisotropy could measure strong scale-dependent bias and stochasticity of the tracers that may not be understood well in a simple parametric model. Proposed intensity mapping experiments \citep{2017arXiv170909066K} have numerous frequency channels and should permit a unique measurement of the clustering of line and continuum radiation as a function of redshift.

\section{Acknowledgements}
\label{sec:ack}
We acknowledge David Spergel for recommendations for cleaning correlated foregrounds.

\appendix

\section{The multi-tracer approach}
\label{apx:multi-tracer}

Recent work \citep{2015ApJ...803...21B, 2015PhRvD..92f3525A, 2015ApJ...812L..22F, 2017ApJ...838...82S, 2017arXiv170909066K} argues that the multi-tracer approach \citep{2011MNRAS.416.3009B, 2009JCAP...10..007M, 2009PhRvL.102b1302S} can apply fruitfully to intensity mapping, and allow a determination of line amplitudes without cosmic variance. This appendix describes this approach in map and power spectrum space \citep{2013MNRAS.436.3089B, 2013MNRAS.432..318A} without foregrounds.

Following the definitions of Section\,\ref{sec:sensitivity}, a model of the observed galaxy redshift survey and intensity maps, respectively is ${\bf x}_{g} = b_g {\boldsymbol \delta} + {\bf n}_g$ and ${\bf x}_{I} = S_L {\boldsymbol \delta} + {\bf n}_I$. We will start with the more standard derivation of the Fisher matrix in map space (e.g. \citet{2011MNRAS.416.3009B}). Per $\ell$ mode, the covariance of the galaxy redshift survey $x_{g, \ell}$ and intensity map $x_{I, \ell}$ is 
\begin{equation}
{\boldsymbol \Sigma}_{gI} = \left ( \begin{matrix} b_g^2 C^{\delta \delta}_\ell + N_{g,\ell} & b_g S_L C^{\delta \delta}_\ell \\ b_g S_L C^{\delta \delta}_\ell & S_L^2 C^{\delta \delta}_\ell + N_{I,\ell} \end{matrix} \right )
\label{eq:bccovparam}
\end{equation}
The Fisher matrix element for determination of covariance parameters $\theta_i$ and $\theta_j$ given the observables $\{ {\bf x}_{g}, {\bf x}_{I} \}$ is then \citep{1997ApJ...480...22T}
\begin{equation}
F_{\theta_i, \theta_j} = \frac{1}{2} Tr \left ( {\boldsymbol \Sigma}_{gI}^{-1} \frac{d{\boldsymbol \Sigma}_{gI}}{d\theta_i} {\boldsymbol \Sigma}_{gI}^{-1} \frac{d{\boldsymbol \Sigma}_{gI}}{d\theta_j} \right ).
\end{equation}
In this setup, the parameters to constrain are ${\boldsymbol \theta} = \{ S_L, C^{\delta \delta}_\ell, N_A \}$. From the inverse of the Fisher matrix, the minimum variance of determination of $S_L$ for each multipole (suppressing the $\ell$ label), marginalized over $C^{\delta \delta}_\ell$ and $N_I$ yields
\begin{equation}
\sigma^2_{S_L} = \frac{1}{v_\ell} \left ( \frac{N_I}{C^{\delta \delta}_\ell} + \frac{N_g}{C^{\delta \delta}_\ell} \frac{S_L^2}{b_g^2} + \frac{N_I N_g}{(C^{\delta \delta}_\ell)^2 b_g^2} + 2 \frac{N_g^2}{(C^{\delta \delta}_\ell)^2} \frac{S_L^2}{b_g^4} \right ),
\end{equation}
where $v_\ell$ is the number of modes in the measurement, which is roughly $(2 \ell + 1) f_{\rm sky}$ for a 2D map on the fraction $f_{\rm sky}$ of the sky. This uses no expansion in the approximation of small noise (e.g. \citet{2011MNRAS.416.3009B}).

Sample variance produces a term linear in $S_L^2$, but here all factors of $S_L$ appear with orders of $N_g / C^{\delta \delta}_\ell$. To simplify expressions throughout to provide some analytic insight, we will assume that galaxy survey shot noise is negligible, or $N_g / C^{\delta \delta}_\ell \ll 1$ (taking no terms beyond order 0). In this case, $\sigma^2_{S_L} = N_I / (v_\ell C^{\delta \delta}_\ell)$. This has the simple interpretation as the noise-to-signal ratio per mode, divided by the number of modes. 

The standard discussion above aligns well with a likelihood analysis of pixel space. Pixel-space likelihoods are expensive in practical data analysis, requiring an $N_{\rm pix} \times N_{\rm pix}$ matrix inverse for $N_{\rm pix}$ pixels. We instead prefer to formulate the analytic Fisher matrix in power spectrum space, where the likelihood for parameters in Section\,\ref{sec:mwpxb} is performed. Working with a likelihood of two-point information also allows more flexibility in the parameter constraint. For example, a likelihood using Eq.\,\ref{eq:bccovparam} has access to information from both the cross-power $b_g S_L C^{\delta \delta}_\ell$ and the auto-power $S_L^2 C^{\delta \delta}_\ell + N_{I}$ of the intensity survey. In practice, the auto-power may not be usable because of additive bias from residual foregrounds, but it can simply be excluded from the two-point likelihood. 

Here the two-point functions are
\begin{eqnarray}
C^{II}_\ell &=& S_L^2 C^{\delta \delta}_\ell + N_I \nonumber \\
C^{gI}_\ell &=& b_g S_L C^{\delta \delta}_\ell \nonumber \\
C^{gg}_\ell &=& b_g^2 C^{\delta \delta}_\ell + N_g.
\end{eqnarray}
Let the observable be the vector ${\boldsymbol \mu} = \{ C^{II}_\ell, C^{gI}_\ell, C^{gg}_\ell \}$ and estimate the parameters $\{ S_L, C^{\delta \delta}_\ell, N_I \}$. In this case, the Fisher matrix for determination of the mean parameters \citep{1997ApJ...480...22T} is
\begin{equation}
F_{\theta_i, \theta_j} = \frac{1}{2} Tr \left [{\rm Cov}({\boldsymbol \mu})^{-1} \left (\frac{d{\boldsymbol \mu}}{d\theta_i} \frac{d{\boldsymbol \mu}^T}{d\theta_j} + \frac{d{\boldsymbol \mu}}{d\theta_j} \frac{d{\boldsymbol \mu}^T}{d\theta_i} \right ) \right ],
\end{equation}
where ${\rm Cov}({\boldsymbol \mu})$ is the covariance of the two-point functions in the observation vector ${\boldsymbol \mu}$. The matrix elements of this covariance matrix are given by Eq.\,\ref{eqn:simplecov}.

Marginalizing over $\{C^{\delta \delta}_\ell, N_I \}$, and ignoring galaxy redshift survey shot noise, one recovers the estimate of $S_L$ with variance $\sigma^2_{S_L} = N_I / (v_\ell C^{\delta \delta}_\ell)$, as above. Sample variance is absorbed in $C^{\delta \delta}_\ell$ as the explanatory variable, allowing $S_L$ to be determined independently. Indeed, when the observable is only the cross-power ${\boldsymbol \mu} = \{ C^{gI}_\ell \}$ and the line amplitude $S_L$ is the only parameter, the variance $\sigma^2_{S_L} = (2S_L^2 + N_I / C^{\delta \delta}_\ell) / v_\ell$ has a term $\propto S_L^2$. 

To simplify the analytics throughout and avoid use of the intensity auto-power, we reduce the observables to the cross-powers and the galaxy redshift survey auto-power, or ${\boldsymbol \mu} = \{ C^{gI}_\ell, C^{gg}_\ell \}$ here. This provides two equations allowing a fit of two parameters $\{ S_L, C^{\delta \delta}_\ell \}$, so assumes that $N_I$ is fixed and known. In this case, the variance on $S_L$, marginalized over $C^{\delta \delta}_\ell$ is again $\sigma^2_{S_L} = N_I / (v_\ell C^{\delta \delta}_\ell)$. This setup is a starting point for Section\,\ref{sec:sensitivity}.

Both approaches above assume that the galaxy redshift survey and line emission are perfect tracers of the same overdensity. In practice, there is some stochasticity $r$ between the two tracers, yielding $C^{gI}_\ell = r b_g S_L C^{\delta \delta}_\ell$. In either the map or power-spectrum space approach, and assuming negligible shot noise, the line amplitude constraint with stochasticity between the tracers becomes
\begin{equation}
\sigma^2_{S_L} = \frac{N_I}{r^2 v_\ell C^{\delta \delta}_\ell} + \frac{S^2_L}{v_\ell} \left ( \frac{1}{r^2} - 1 \right ).
\end{equation}
Note that this does contain cosmic variance terms $\propto S^2_L$ when $r \neq 1$. When there is stochasticity between the populations, the component of variance that is not correlated between the tracers adds cosmic variance. Furthermore, the thermal noise term $N_I/(r^2 v_\ell C^{\delta \delta}_\ell)$ scales as $1/r^2$. On the large angular scales considered in Section\,\ref{sec:mwpxb}, the galaxy redshift survey and intensity trace the same underlying linear perturbations, where $r \approx 1$ is a reasonable approximation.

Evasion of cosmic variance in the line amplitude estimate can be understood in both map and power spectrum space. In map space, the galaxy redshift survey provides a spatial template of overdensity whose amplitude can be fit in the intensity survey without cosmic variance (it is a determination of a mean amplitude rather than of variance). In power spectrum space, cosmic variance is correlated between the intensity and galaxy redshift survey. Fitting jointly for $C^{\delta \delta}_\ell$ and $S_L$ makes $C^{\delta \delta}_\ell$ the explanatory variable of the cosmic covariance. $S_L$ only appears as an amplitude in the intensity survey so is determined independently of the cosmic covariance. 

\section{A multiband model with separable foregrounds}
\label{app:sepfg}

Extend the model above to two intensity bands, $A$ and $B$ at frequencies $\nu_A$ and $\nu_B$. Let the amplitude of correlated signal in maps $A$ and $B$ be $\phi$ and $\psi$. Add foreground covariance with spatial anisotropy $C^{uu}_\ell$ and amplitude $f(\nu)$ to both intensity bands. The two-point functions are
\begin{eqnarray}
C^{AA}_\ell &=& \phi^2 C^{\delta \delta}_\ell + f(\nu_A)^2 C^{uu}_\ell + N_A \nonumber \\
C^{BB}_\ell &=& \psi^2 C^{\delta \delta}_\ell + f(\nu_B)^2 C^{uu}_\ell + N_B\nonumber \\
C^{AB}_\ell &=& \phi \psi C^{\delta \delta}_\ell + f(\nu_A) f(\nu_B) C^{uu}_\ell \nonumber \\
C^{Ag}_\ell &=& \phi b_g C^{\delta \delta}_\ell \nonumber \\
C^{Bg}_\ell &=& \psi b_g C^{\delta \delta}_\ell \nonumber \\
C^{gg}_\ell &=& b_g^2 C^{\delta \delta}_\ell + N_g.
\end{eqnarray}
In the limit of negligible shot noise, the covariance of determinations $\hat \phi$ and $\hat \psi$ is
\begin{eqnarray}
{\rm Cov}(\hat \phi, \hat \phi) &=& \frac{f(\nu_A)^2 C^{uu}_\ell + N_A}{v_\ell C^{\delta \delta}_\ell} \nonumber \\
{\rm Cov}(\hat \phi, \hat \psi) &=& \frac{f(\nu_A) f(\nu_B) C^{uu}_\ell}{v_\ell C^{\delta \delta}_\ell} \nonumber \\
{\rm Cov}(\hat \psi, \hat \psi) &=& \frac{f(\nu_B)^2 C^{uu}_\ell + N_B}{v_\ell C^{\delta \delta}_\ell}.
\end{eqnarray}
Extending this to many bands gives covariance the form
\begin{equation}
{\bf N} = \frac{1}{v_\ell C^{\delta \delta}_\ell} ({\bf N}_{\rm th} + C^{uu}_\ell {\bf f} {\bf f}^T),
\label{eq:rankonefg}
\end{equation}
where ${\bf N}_{\rm th}$ is the diagonal matrix of thermal noise in each band ($N_A$ and $N_B$ above) and ${\bf f}_i = f(\nu_i)$ is the foreground spectrum evaluated in each of the bands. In summary, if the Milky Way adds common spatial structure to the intensity maps, it will appear as a rank-1 term in the bandpower covariance as a function of frequency.

\section{Multiband SED reconstruction}
\label{app:multiband}

Here we develop a simple linear model for an intensity survey with many channels. Let the multiband intensity mapping survey have frequencies labeled $\nu_i$ and find the cross-correlation with the galaxy redshift survey at fixed redshift $z$. Following notation in Section\,\ref{sec:sensitivity}, model this power spectrum as
\begin{equation}
C_\ell[{\bf x}_I(\nu_i) \times {\bf x}_g(z)]
= [S_L \xi(\nu_i) + S_C(\nu_i)]\cdot b_g C^{\delta \delta}_\ell(z)
\label{eqn:multibandm1}
\end{equation}
In the case of Planck, quasars in \zrange\ only correlate with one frequency channel, so $\xi(\nu_i) = 1$ in $\nu_i=545$\,GHz and $\xi(\nu_i)=0$ at the other Planck frequencies. In a more typical intensity survey with narrow spectral channels, the clustering of large scale structure at low-$k_\parallel$ produces correlations between channels in the kernel $\xi(\nu_i)$ \citep{2013JCAP...11..044D, 2011PhRvD..84d3516C}.

In the illustrative limit of negligible shot noise, the galaxy redshift survey auto-power fixes $C^{\delta \delta}_\ell(z)$, including cosmic variance, making the model for the cross-powers in Eq.\,\ref{eqn:multibandm1} approximately linear in the amplitude $[S_L \xi(\nu_i) + S_C(\nu_i)]$. We can therefore use a simple analytic model of the cross-power as a function of frequency to understand the joint constraint of correlated line and continuum emission. In practice with BOSS quasars, shot noise is not negligible, but the linear approach gives intuition for the geometry of the joint analysis of correlated and uncorrelated continuum. Future intensity mapping experiments would be complemented well by galaxy redshift surveys with small shot noise on the scales of interest.

Form a new data vector ${\bf d}_\ell$ from the measured cross-powers at frequency $\nu_i$ divided by $b_g \hat C^{\delta \delta}_\ell(z)$ inferred from the galaxy redshift survey
\begin{equation}
d^\times_\ell(\nu_i) = \hat C_\ell[{\bf x}_I(\nu_i) \times {\bf x}_g(z)] / (b_g \hat C^{\delta \delta}_\ell(z)).
\label{eqn:datavec}
\end{equation}
This is an estimator for $S_L \xi(\nu_i) + S_C(\nu_i)$. Model the SED of the correlated continuum as the sum of linear components $g_j(\nu_i)$ and amplitudes $S_{C,j}$,
\begin{equation}
S_C(\nu_i) = \sum_j^{N_{\rm comp}} S_{C,j} g_j(\nu_i).
\end{equation}
Pack each spectral component into a vector ${\bf g}_j|_i = g_j(\nu_i)$, and pack the signal correlation kernel into a vector ${\boldsymbol \xi}|i = \xi(\nu_i)$. In the case of widely separated bins in the Planck analysis, this is a delta function at 545\,GHz. Pack the line signal and $N_{\rm comp}$ correlated continuum modes into ${\bf M} \equiv [{\boldsymbol \xi}, {\bf g}_1, ... {\bf g}_{N_{\rm comp}}]$. The linear amplitude parameter vector is then ${\boldsymbol \theta} \equiv [ S_L, {\boldsymbol S}_C ]$ where ${\boldsymbol S}_C$ are the $S_{C,j}$ linear parameters for the SED. With these substitutions, the model for multiband cross-powers in Eq.\,\ref{eqn:multibandm1} is 
\begin{equation}
{\bf d}^\times_\ell = {\bf M} {\boldsymbol \theta} + {\bf n}_\times,
\label{eqn:multiband}
\end{equation}
where ${\bf n}_\times$ is the noise of the cross-power measurements drawn from covariance ${\bf N}_\times$ that includes both thermal noise (${\bf N}_{\rm th}$) and foreground (${\bf N}_{\rm fg}$) covariance as 
\begin{equation}
{\bf N}_\times = \frac{1}{\nu_\ell C^{\delta \delta}_\ell} ({\bf N}_{\rm th} + {\bf N}_{\rm fg}).
\end{equation}
(Note that the data vector defined in Eq.\,\ref{eqn:datavec} in an estimator for the amplitudes for $\phi$ and $\psi$ in App.\,\ref{app:sepfg}, but applied to many bands. This results in the same form for the covariance as Eq.\,\ref{eq:rankonefg}).

This model solves jointly for the line amplitude $S_L$ and spectral baseline amplitudes ${\boldsymbol S}_C$ that describe the correlated continuum emission \citep{2017ApJ...838...82S}. The linear estimate $\hat {\boldsymbol \theta}$ and its covariance ${\boldsymbol \Sigma}_{\boldsymbol \theta}$ are 
\begin{eqnarray}
\hat {\boldsymbol \theta} &=& {\boldsymbol \Sigma}_{\boldsymbol \theta} {\bf M}^T {\bf N}_\times^{-1} {\bf d}_\ell \nonumber \\
{\boldsymbol \Sigma}_{\boldsymbol \theta} &=& ({\bf M}^T {\bf N}_\times^{-1} {\bf M})^{-1}.
\label{eqn:linearmodel}
\end{eqnarray}
We have written this expression for a single multipole $\ell$ for simplicity, but the model could be extended as a likelihood on the cross-powers at all $\ell$ in practice.

This linear form recovers our primary results so far. Eq.\,\ref{eqn:linearmodel} recovers Eq.\,\ref{eqn:cleanedsl} in the case where ${\bf M} = [[1,0]^T]$ (only $S_L$ is fit from the data) and $r_F=1$. Here, ${\bf N}_{\rm fg} = C^{uu}_\ell [f(\nu_I), f(\nu_V)]^T [f(\nu_I), f(\nu_V)]$ and ${\bf N}_{\rm th} = {\rm diag}[N_I, N_V]$. Adding one correlated SED component as ${\bf M} = [[1,0]^T, [g(\nu_I), g(\nu_V)]^T]$ recovers Eq.\,\ref{eqn:totalerr}, again with $r_F=1$. 

\section{Geometry of the joint correlated and uncorrelated continuum estimation}
\label{app:lowrankgeo}

We can now consider the effect of Milky Way foregrounds by letting the covariance ${\bf N}_\times^{-1}$ be the rank-1 foreground model Eq.\,\ref{eq:rankonefg}, equivalent to a spatial pattern common across all frequencies. In practice, Milky Way foregrounds will have higher rank, which can be included in ${\bf N}_\times^{-1}$. In this simplified setting, we can continue analytically by applying
the Sherman-Morrison formula twice to give
\begin{eqnarray}
{\boldsymbol \Sigma}_{\boldsymbol \theta} &=& \frac{1}{v_\ell C^{\delta \delta}_\ell} \left ( {\boldsymbol \Sigma}_{\rm th} +\frac{C^{uu}_\ell {\boldsymbol \alpha}_f {\boldsymbol \alpha}_f^T}{1 + C^{uu}_\ell \Delta} \right ) 
\label{eqn:multiband_rankone}
\end{eqnarray}
where 
\begin{eqnarray}
\Delta &\equiv& {\bf f}^T {\bf N}_{\rm th}^{-1}({\bf f} - {\bf M} {\boldsymbol \alpha}_f) \nonumber \\ 
{\boldsymbol \Sigma}_{\rm th} &\equiv& ({\bf M}^T {\bf N}_{\rm th}^{-1} {\bf M})^{-1} \nonumber \\
{\boldsymbol \alpha}_f &\equiv& {\boldsymbol \Sigma}_{\rm th} {\bf M}^T {\bf N}_{\rm th}^{-1} {\bf f}.
\label{eqn:multiband_defs}
\end{eqnarray}

Here, ${\boldsymbol \Sigma}_{\rm th}$ is the variance for the estimate of ${\boldsymbol \theta}$ with only thermal noise and no foregrounds. ${\boldsymbol \alpha}_f$ is the optimal linear estimator, with thermal noise only, applied to the foreground SED. Hence $C^{uu}_\ell {\boldsymbol \alpha}_f {\boldsymbol \alpha}_f^T$ is the amplitude of the foregrounds in the ${\boldsymbol \theta}$ parameter basis. Finally, $\Delta$ is a discriminant that describes how much information about the uncorrelated foregrounds is left after fitting for the correlated spectral components in ${\bf M}$. 

We can now consider the geometry of the joint correlated and uncorrelated continuum separation problem. The key quantity here is the discriminant $\Delta$. This is a measure of the remaining RMS of the uncorrelated foregrounds ${\bf f}$ after fitting out the correlated continuum modes ${\bf g}_i$. (To see this, note that ${\boldsymbol \alpha}_f$ projects ${\bf f}$ onto the parameter basis ${\boldsymbol \theta}$ and then ${\bf M}$ projects onto a spectrum differenced with ${\bf f}$ in $\Delta$.) 

If ${\bf M}$ is a complete spectral basis, then $\Delta = 0$ regardless of the spectral shape of ${\bf f}$. This will be the case any time the number of independent parameters in the correlated SED model is equal to the number of bands, such as in the case of Planck cross BOSS. In this regime,
\begin{equation}
{\boldsymbol \Sigma}_{\boldsymbol \theta} = \frac{1}{v_\ell C^{\delta \delta}_\ell} \left ( {\boldsymbol \Sigma}_{\rm th} + C^{uu}_\ell {\boldsymbol \alpha}_f {\boldsymbol \alpha}_f^T \right ).
\label{eqn:multibandsimp}
\end{equation}
These two terms correspond to the thermal noise and foreground components in Eq.\,\ref{eqn:totalerr} with $r_F=1$. The second term represents foregrounds that contaminate the line amplitude estimate after marginalizing the correlated continuum spectral modes ${\bf g}_j$. An observation with few bands may not provide sufficient degrees of freedom to deweight uncorrelated continuum foregrounds and model the correlated continuum foreground. 

Adding a template leads to a qualitative improvement in the foreground cleaning. When $N_{\rm band} = N_{\rm param}$, Eq.\,\ref{eqn:multibandsimp} shows that the line amplitude variance always has a term proportional to $C^{uu}_\ell$. By adding a template, $C^{uu}_\ell \Delta$ in the denominator of Eq.\,\ref{eqn:multiband_rankone} can remain large. In the limit of large foregrounds, 
\begin{equation}
{\boldsymbol \Sigma}_{\boldsymbol \theta} = \frac{1}{v_\ell C^{\delta \delta}_\ell} \left ( {\boldsymbol \Sigma}_{\rm th} + \frac{{\boldsymbol \alpha}_f {\boldsymbol \alpha}_f^T}{\Delta} \right ).
\end{equation}
Now the foreground brightness $C^{uu}_\ell$ drops out, and the numerator ${\boldsymbol \alpha}_f {\boldsymbol \alpha}_f^T$ is suppressed by $\Delta$. Again in this limit, the variance of the $S_L$ estimate is not compromised by bright foregrounds, making this a promising approach. This is the extension of Eq.\,\ref{eqn:totalwtemplate} to multiple bands.

\section{The impact of stochasticity between Milky Way tracers and dust emission}
\label{sec:fishermw}

Extend the template model of Eq.\,\ref{eqn:totalmodelwithtemplate} to allow for imperfect correlation $r_M = C^{Mu}_\ell / \sqrt{C^{MM}_\ell C^{uu}_\ell}$ with the actual foreground ${\bf x}_u$ in the intensity map. The correlations between this map and itself ($MM$), the intensity map ($MI$), and veto map ($MV$) are
\begin{eqnarray}
C^{MM}_\ell &=& A_M^2 C^{uu}_\ell + N_M \\
C^{MI}_\ell &=& r_M A_M f(\nu_I) C^{uu}_\ell \\
C^{MV}_\ell &=& r_M A_M f(\nu_V) C^{uu}_\ell.
\end{eqnarray}
For simplicity, let $N_M \rightarrow 0$, so the template has no noise but is still an imperfect tracer of the true foreground. Then,
\begin{eqnarray}
\sigma^2_{S_L} &=& \frac{C^{uu}_\ell}{v_\ell C^{\delta \delta}_\ell}  \left [f(\nu_I) - f(\nu_V) \frac{g(\nu_I)}{g(\nu_V)} \right ]^2 (1 - r_M^2) \nonumber \\
&&+ \frac{1}{v_\ell C^{\delta \delta}_\ell} \left [N_I + \left [ \frac{g(\nu_I)}{g(\nu_V)} \right ]^2 N_V \right ].
\end{eqnarray}

\section{Redshift-space distortions on the Limber-approximated angular power spectrum}
\label{sec:rsdlimber}

In the paper we argue that redshift-space distortions can be neglected for the bulk of our analysis.  To prove this, we first derive the angular power spectrum, which is directly related to the CIB-LSS angular cross-power spectrum we use in our analysis, under the Limber approximation \citep{2008PhRvD..78l3506L}.  We begin with the full expression for the angular power spectrum in redshift space, given by $C_\ell=C_\ell^{00}+2\beta C_\ell^{0r}+\beta^2 C_\ell^{rr}$ where $\beta$ is the RSD parameter given by the growth rate-clustering bias ratio,
\begin{eqnarray}\label{E:clrsd}
C_\ell^{00}&=&\frac{2}{\pi}\int dk\,k^2P(k)[W_\ell^0(k)]^2\nonumber\\
C_\ell^{0r}&=&\frac{2}{\pi}\int dk\,k^2P(k)W_\ell^0(k)W_\ell^r(k)\nonumber\\
C_\ell^{rr}&=&\frac{2}{\pi}\int dk\,k^2P(k)[W_\ell^r(k)]^2\, ,
\end{eqnarray}
where $P(k)$ is the matter power spectrum at redshift $z=0$ and the window function $W_\ell(k)=W_\ell^0(k)+W_\ell^r(k)$.  Given the selection function $\phi(r)$, we can write \citep{2007MNRAS.378..852P}
\begin{eqnarray}
W_\ell^0(k)&=&\int dr\,\phi(r)j_\ell(kr)\nonumber\\
W_\ell^r(k)&=&\int dr\,\phi(r)\left[\frac{2\ell^2+2\ell-1}{(2\ell-1)(2\ell+3)}j_\ell(kr)-\frac{\ell(\ell-1)}{(2\ell-1)(2\ell+1)}j_{\ell-2}(kr)-\frac{(\ell+1)(\ell+2)}{(2\ell+1)(2\ell+3)}j_{\ell+2}(kr)\right]\, .
\end{eqnarray}

We wish to perform operations similar to those in \citet{2008PhRvD..78l3506L} to get the Limber approximation for this expression.  Using Eqs. 7-11 in \citet{2008PhRvD..78l3506L}, we can show that the Limber approximation at first-order is
\begin{eqnarray}
\sqrt{\frac{2}{\pi}}\int dr\,\phi(r)j_\ell(kr)&\simeq&\frac{1}{k\sqrt{\ell+1/2}}\phi\left(\frac{\ell+1/2}{k}\right)\nonumber\\
&\simeq&\frac{1}{k\sqrt{\ell}}\phi\left(\frac{\ell}{k}\right)\, ,
\end{eqnarray}
where we approximate $\ell+1/2\simeq\ell$ to make the following equations less cluttered.  Inserting this into Eq.~\ref{E:clrsd}, we can write
\begin{eqnarray}\label{E:clrsdfinal}
C_\ell^{00}&=&\frac{1}{\ell}\int dk\,\phi^2\left(\frac{\ell}{k}\right)P(k)\nonumber\\
C_\ell^{0r}&=&\frac{1}{\sqrt{\ell}}\int dk\,\phi\left(\frac{\ell}{k}\right)F(\ell,k)P(k)\nonumber\\
C_\ell^{rr}&=&\int dk\,F^2(\ell,k)P(k)\, ,
\end{eqnarray}
where
\begin{eqnarray}
F(\ell,k)=\frac{2\ell^2+2\ell-1}{(2\ell-1)(2\ell+3)}\frac{\phi(\ell/k)}{\sqrt{\ell}}-\frac{\ell(\ell-1)}{(2\ell-1)(2\ell+1)}\frac{\phi[(\ell-2)/k]}{\sqrt{\ell-2}}-\frac{(\ell+1)(\ell+2)}{(2\ell+1)(2\ell+3)}\frac{\phi[(\ell+2)/k]}{\sqrt{\ell+2}}\, .
\end{eqnarray}

Now we wish to use this expression to show when RSD can be neglected.  In this argument we will assume $\phi(r)$ is a tophat distribution centered at $r=R$ with full width $\Delta R$.  In this case, it is evident that if $\ell/k-R\ll\Delta R$, then $\phi[(\ell\pm2)/k]\simeq\phi(\ell/k)$.  Additionally assuming $\ell$ is large enough such that $\sqrt{\ell\pm 2}\simeq\sqrt{\ell}$, we can show that in this case $F(\ell,k)\simeq 1/(4\ell^{5/2})$ which decreases rapidly to zero, eliminating the RSD effect.  Taking the approximation that $\ell/k\sim R$, we argue that all the $\phi$s are nonzero when $2/k<\Delta R/2$, which implies the RSD effect vanishes when $\ell\gtrsim 4R/\Delta R$.  For the quasars $4R/\Delta R\sim 17$, while for the LRGs $4R/\Delta R\sim 10$.  Thus, we are justified in neglecting RSD for our analysis.

\bibliographystyle{apj}
\bibliography{main}

\end{document}